\documentclass[fleqn,usenatbib]{mnras}

\usepackage{newtxtext,newtxmath}
\usepackage{amsmath}
\usepackage{subfig}
\usepackage[T1]{fontenc}
\usepackage{xcolor}
\DeclareRobustCommand{\VAN}[3]{#2}
\let\VANthebibliography\thebibliography
\def\thebibliography{\DeclareRobustCommand{\VAN}[3]{##3}\VANthebibliography}

\usepackage{graphicx}	
\usepackage{amsmath}	

\newcommand{\RH}[1]{#1}

\newcommand{\tth}{\textsc{the300}}

\newcommand{\hMpc}{{\ifmmode{\,h^{-1}{\rm Mpc}}\else{$h^{-1}$Mpc}\fi}}
\newcommand{\hkpc}{{\ifmmode{\,h^{-1}{\rm kpc}}\else{$h^{-1}$kpc}\fi}}
\newcommand{\hMsun}{{\ifmmode{\,h^{-1}{\rm {M_{\odot}}}}\else{$h^{-1}{\rm{M_{\odot}}}$}\fi}}
\newcommand{\Msun}{M$_\odot$}
\newcommand{\Mstar}{{\ifmmode{\,M_{*}}\else{$M_{*}$}\fi}}
\newcommand{\Mhalo}{{\ifmmode{\,M_{\rm halo}}\else{$M_{\rm halo}$}\fi}}
\renewcommand{\tilde}[1]{\stackrel{\sim}{\smash{#1}\rule{0pt}{1.1ex}}}
\newcommand{\theth}{\textsc{The Three Hundred\,}}
\newcommand{\ahf}{\textsc{AHF}}
\newcommand{\caesar}{\textsc{Caesar}}

\newcommand{\gadgetx}{\textsc{Gadget-X}}
\newcommand{\simba}{\textsc{Gizmo-Simba}}
\newcommand{\gadgetmusic}{\textsc{Gadget-MUSIC}}

\def\kms{\,{\rm km}\,{\rm s}^{-1}}

\title[Boundless Baryons]{Boundless baryons: how diffuse gas contributes to anisotropic tSZ signal around simulated Three Hundred clusters}

\author[M. Lokken et al.]{Martine Lokken,$^{1,2,3}$\thanks{E-mail: m.lokken@mail.utoronto.ca}
Weiguang Cui,$^{4,5,6}$
J. Richard Bond,$^{1,2}$
Ren\'ee Hlo\v{z}ek,$^{1,3}$ Norman Murray,$^{2}$
\newauthor
Romeel Dav\'e,$^{6,7}$ and Alexander van Engelen$^{8}$
\\
$^{1}$David A. Dunlap Department of Astronomy and Astrophysics, University of Toronto, 50 St. George Street, Toronto, Ontario, M5S 3H4 Canada\\
$^{2}$Canadian Institute for Theoretical Astrophysics, University of Toronto, 60 St. George St., Toronto, ON M5S 3H4, Canada\\
$^{3}$Dunlap Institute of Astronomy \& Astrophysics, 50 St. George St., Toronto, ON M5S 3H4, Canada\\
$^{4}$ Departamento de Física Teórica, M-8, Universidad Autónoma de Madrid, Cantoblanco 28049, Madrid, Spain\\
$^{5}$ Centro de Investigación Avanzada en Física Fundamental (CIAFF), Universidad Aut\'{o}noma de Madrid, Cantoblanco, 28049 Madrid, Spain\\
$^6$ Institute for Astronomy, University of Edinburgh, Royal Observatory, Edinburgh EH9 3HJ, United Kingdom\\
$^7$ Department of Physics and Astronomy, University of the Western Cape, Bellville, Cape Town 7535, South Africa\\
$^8$ 
School of Earth and Space Exploration, 
Arizona State University, Tempe, AZ 85287, USA
}

\date{Accepted XXX. Received YYY; in original form ZZZ}

\pubyear{2023}

\begin{document}
\label{firstpage}
\pagerange{\pageref{firstpage}--\pageref{lastpage}}
\maketitle

\begin{abstract}
Upcoming advances in galaxy surveys and cosmic microwave background data will enable measurements of the anisotropic distribution of diffuse gas in filaments and superclusters at redshift $z=1$ and beyond, observed through the thermal Sunyaev-Zel'dovich (tSZ) effect. These measurements will help distinguish between different astrophysical feedback models, account for baryons that appear to be `missing' from the cosmic census, and present opportunities for using locally-anisotropic tSZ statistics as cosmological probes. This study seeks to guide such future measurements by analysing whether diffuse intergalactic gas is a major contributor to anisotropic tSZ signal in \theth\ \simba\, hydrodynamic simulations. We apply multiple different halo boundary and temperature criteria to divide concentrated from diffuse gas at $z=1$, then create mock Compton-$y$ maps for the separated components. The maps from 98 simulation snapshots are centred on massive galaxy clusters, oriented by the most prominent galaxy filament axis, and stacked. Results vary significantly depending on the definition used for diffuse gas, indicating that assumptions should be clearly defined when claiming observations of the warm-hot intergalactic medium. In all cases, the diffuse gas is important, contributing 25--60\% of the tSZ signal in the far field ($>4 h^{-1}$ comoving Mpc) from the stacked clusters. The gas 1--2 virial radii from halo centres is especially key. Oriented stacking and environmental selections help to amplify the signal from the warm-hot intergalactic medium, which is aligned but less concentrated along the filament axis than the hot halo gas.
\end{abstract}




\section{Introduction}
The distribution and state of baryons in the cosmic web has become an increasingly important puzzle in cosmology and astrophysics. The fractional contribution of baryons to the total energy budget of the Universe today, $\Omega_\mathrm{b}$, can be predicted from early-Universe cosmic microwave background (CMB) data assuming a cosmological model, such as the current concordance model $\Lambda$CDM \citep[e.g.,][]{WMAP2013, Planck2018, ACT2020}. A full census of baryons via direct detection in the late-time Universe would be an excellent test of the cosmological model. However, for years, the estimated baryon fraction from low-redshift observations fell far below the prediction from the early Universe, motivating a search for the `missing baryons' \citep{CenOstriker1999}.

Theory paints the following picture of the history of baryons. As the Universe evolved from its initial nearly-uniform state, baryons followed the dark matter through linear and then non-linear structure growth, populating the filamentary cosmic web and evacuating void regions. Intergalactic hydrogen at $z\gtrsim2$ can be observed via the Lyman-$\alpha$ forest, and with these observations the predicted baryons are largely accounted for. However, the intergalactic hydrogen became increasingly ionized and shock-heated by structure formation, while also becoming more diffuse due to cosmological expansion. Additionally, active galactic nuclei (AGN) and supernovae (SN) began to eject and re-heat gas that had previously fallen into galaxies and cooled. The combination of low densities ($n_b\sim10^{-6}-10^{-5}$ cm$^{-3}$), intermediate temperatures ($10^5-10^7$ K), and ionization make such gas challenging to observe, hence the moniker of `missing' baryons. By $z=0$, simulations predict that $\sim30-70\%$ of the baryonic mass exists in this so-called warm-hot intergalactic medium, or WHIM \citep{CenOstriker1999, Dave2001, CenOstriker2006, Penton2004, Shull2012, Nicastro2008, McQuinn2016, Cui2019, Sorini2022}.

The wide range in these predictions is due to various evolutions of the WHIM which occur depending on the implemented feedback model \citep{Dave2001}. Constraining the evolution of the WHIM is important not only for the cosmological energy census, but also for its potential as a tracer of low-redshift cosmic web structure, a playground for testing the behavior of dark energy and dark matter. The sensitivity of the WHIM to galactic feedback processes means it can also be used to constrain the physics of AGN and SN, an area of high interest for cosmology--e.g., the redistribution of baryons from astrophysical feedback is considered to be one of the most important unknowns for utilizing the full power of current and future lensing surveys \citep{Harnois-Deraps2015}. While feedback presents a nuisance for cosmology, it is interesting for enhancing our understanding of black holes and stellar evolution \citep{Kormendy2013,Yuan2014, Somerville2015}. The states of the circumgalactic and intergalactic media are of interest in their own right \citep[see][for reviews]{Meiksin2009, Tumlinson2017}, and have important relationships with galaxy evolution \citep{Oppenheimer2008, Dave2011}.

The WHIM induces very weak signals in maps of the CMB through the Sunyaev-Zel'dovich (SZ) effects \citep{SZ1972}. The thermal Sunyaev-Zel'dovich effect is sensitive to gas pressure along the line-of-sight; it is parameterized by the Compton-$y$ parameter,
\begin{equation}\label{eq:tSZ}
    y =  \int d\ell n_\mathrm{e} \frac{k_\mathrm{B} T_\mathrm{e}}{m_\mathrm{e} c^2} \sigma_\mathrm{T}.
\end{equation}
$y$ is a line-of-sight integral of the number density of electrons $n_\mathrm{e}$ and the electron temperature $T_\mathrm{e}$. $\sigma_\mathrm{T}$ is the Thomson cross section and $m_\mathrm{e}$ is the electron mass. Due to the dependence on both density and temperature, signals from galaxy clusters are orders-of-magnitude higher than signals from intergalactic filaments, which are well below the noise in current maps. Nevertheless, maps have recently become sensitive enough for significant detection of tSZ signals from an intermediate regime: a cluster pre-merger bridge \citep{Planck2013,Bonjean2018, Hincks2022}. Meanwhile, the kinematic SZ effect, depending on gas density and bulk velocity, is significantly smaller than tSZ and more difficult to reconstruct. However, recent advances \citep{Hill2016, Ferraro2016, Kusiak2021, Schaan2021} have forwarded the kSZ as a promising WHIM probe, especially in combination with tSZ measurements. We will focus only on tSZ in this work.

Several studies have developed new stacking techniques for teasing out the weak tSZ signals from filaments using combinations of CMB and galaxy survey data. Stacking averages together $N$ images of the same or similar objects, which contain signal in the same location but have varying noise, to create a single image in which the signal-to-noise ratio (SNR) is enhanced by $\sqrt{N}$. By stacking a $y$ map from \textit{Planck} satellite data on locations of pairs of galaxies, \citet{degraaff2019} and \citet{Tanimura2019} found evidence for tSZ signal from a stacked intercluster filament. The former group posited that $\sim80\%$ percent of the $y=(0.6\pm0.2)\times10^{-8}$
filament signal came from gas beyond dark matter halos. \citet[][hereafter L22]{Lokken2022} used a distinct oriented stacking method to measure the large-scale anisotropy of tSZ around clusters in Atacama Cosmology Telescope (ACT) data, detecting a signal from the combined emission of neighboring halos and filament gas at the 3.5$\sigma$ level and demonstrating that the signal is stronger in supercluster regions, as defined by galaxy data.

These methods have been applied only to data from $z<1$, where there is sufficient galaxy number density in the current state-of-the-art large galaxy surveys \citep[e.g., the Sloan Digital Sky Survey (SDSS) and Dark Energy Survey (DES),][] {Bolton2012,Pandey2021, Porredon2021}. All oriented stacking methods will improve with higher galaxy number density. As upcoming surveys map the cosmic web of galaxies to higher redshifts with greater number density and precision \citep[e.g., the Vera C. Rubin Observatory and the Dark Energy Spectroscopic Instrument,][]{LSST2019, DESI2016}, and anisotropic statistical methods improve, it may finally be possible to observationally constrain the WHIM evolution from $z\sim1$ to $z=0$.

To motivate such studies, in this work we use oriented stacking to examine tSZ signals from the diffuse gas in hydrodynamic simulations at $z=1$. Following the applications to both simulated and observed data in L22, which determined the orientation of filaments and superclusters on scales of $\sim10-20$ comoving Mpc, we conduct a similar study on the smaller end of this scale range using the \simba\ run of \theth\ project\footnote{\url{https://the300-project.org/}, part of the data are available through this website: \url{https://the300.ft.uam.es/}, upon request.} \citep[\tth\ for short,][]{Cui2018}. In brief, these simulations are run with \textsc{Simba} hydrodynamics \citep{Dave2019}, include both SN and AGN feedback, and re-simulate reasonably large (15~\hMpc\ comoving) regions surrounding 324 different massive clusters with a competitive resolution ($\sim10^8$~\hMsun). Thus, \tth\ runs provide an excellent option for studying the filamentary structure around clusters due to the sufficient numbers of clusters for stacking (competitive or better than the numbers formed in large cosmological simulations) and the larger extent than typical cluster zoom simulations. Furthermore, their nature as zoom simulations makes it feasible for future work to re-simulate selected regions with different feedback models. We focus our study at $z=1$, the new frontier of galaxy-CMB cross-correlation studies, where the proto-cluster regions are still coalescing along complex filamentary structures. The detailed nature of the simulations enables us to address a key question: \textit{how much tSZ signal does diffuse gas contribute to filaments?}

We begin by dividing the particles into diffuse and concentrated categories by several methods. The fiducial method applies a threshold at the halo virial radius $R_{200c}$ (where the density enclosed is 200 times the critical density), considering the particles within this radius as halo-associated and those beyond as diffuse. Compton-$y$ maps are then created for the different particle sets, oriented using information from the galaxy field, and stacked. We examine the stacks to determine the fractional contribution to the $y$ signal from halo versus diffuse gas.

There are two main motivations for dividing the halo from the diffuse gas in the context of oriented SZ studies. First, the tSZ effect has the potential to be a probe of cosmological structure beyond 2-point statistics. With advances in oriented stacking, the tSZ can start to provide information about the cosmic web shape and structure beyond the densest, highest-pressure nodes. However, a cosmological probe is only as good as its respective modelling. To be able to simulate the effects of varied cosmologies, the best approach is to use analytic theory, N-body, or rapid predictive simulations for the dark matter, and post-process to apply a prescription for the baryons \citep[e.g., as done in][]{Stein2019, Pandey2019, Raghunathan2022}. The alternative -- many large-volume hydrodynamic simulations with varied cosmologies -- is currently computationally infeasible, although it is being tested in smaller volumes by the \textsc{Camels} project \citep{CAMELS2021}. The leading prescriptions for tSZ are based on the halo model, assigning baryons to halo locations with a pressure profile which is well-understood to some radius. However, it is unclear whether this prescription sufficiently captures the tSZ signal, especially for locally-anisotropic studies such as oriented stacking, due to its lack of description for the diffuse gas beyond halos. Our study asks: \textit{is it possible to capture the majority of the extended tSZ signal by using halo modeling? If so how far, and at what halo masses, need to be accurately modeled?} This will guide future oriented SZ studies: if the unassociated gas is unimportant, then the focus should remain on perfecting the halo pressure profile models in order to be able to use the tSZ to probe beyond-two-point statistics. If it is important, more attention should be devoted to modelling the field component of the tSZ.

The second major motivation is to guide studies which seek to find the `missing' baryons. For the observational analyses that aim to do this with tSZ, the contribution from known baryons must be subtracted: e.g., the $y$ map should be masked at halo positions out to some radius, or the contribution from halos should be modelled (as done in \citealt{degraaff2019, Tanimura2019}). Thus it is useful to know what fraction of $y$ is captured by halo models and what residual is expected in the current best hydrodynamical simulations.

The answers to these questions depend on the implemented feedback model and thus will be specific to \tth\ feedback mechanisms. Analyzing how the diffuse $y$ signal changes for different feedback prescriptions is an interesting and important question for future work. Nevertheless, it is still useful to explore the questions given the particular setup of \tth, which is one of the leading simulations currently available with enough volume for studies of extended tSZ. Even an approximate estimate for the importance of diffuse gas in tSZ studies will help motivate analysis with upcoming tSZ and galaxy survey data.

Throughout the paper, we use the cosmological parameters from the \citet{Planck2016}, also used in \tth\ simulations. All quoted distances are in comoving  Mpc. Sec.~\ref{sec:sim_data} describes details of the simulations. Sec.~\ref{sec:methods} explains the methods including selection of a central halo for the stacks, reduction of the snapshot sample, splitting of the halo-associated and diffuse gas, creation of $y$ maps, and orientation of the maps using galaxy data. In Sec.~\ref{sec:results}, we present the stacked maps and analyse their radial profiles in a multipole decomposition. We also discuss the association of gas with halos and compare the filament to off-filament axes. In Sec.~\ref{sec:superclustering}, $y$ maps are sub-selected based on measures of density and elongation, and the resulting differences in signal are analyzed. Sec.~\ref{sec:conclusions} presents discussions and conclusions.




\section{Simulation data}\label{sec:sim_data}
The cluster samples used in this study are coming from \tth\ \simba\ simulations. \tth\ selects the most massive galaxy clusters from the MultiDark simulation \citep[MDPL2,][]{Klypin2016}, \RH{which is a series of simulations that cover a range of masses from $10^{10}M_\odot - 10^{15}M_\odot$ and volumes up to $50~\mathrm{Gpc}^3$.} The simulation re-generates the initial conditions using the {\sc Ginnungagap} code\footnote{Code available online: \url{https://github.com/ginnungagapgroup/ginnungagap}} with a zoomed-in technique, which allows us to only simulate the cluster regions with different hydrodynamic codes. The zoom regions are extended to 15~\hMpc\ from the centres of these clusters and cover over $5\times R_{200c}$, which allows us to investigate the environments around the clusters. Many studies, such as \cite{Haggar2020,Kuchner2020,Rost2021,Kuchner2021}, benefit from this large re-simulation region. Since MDPL2 adopted the Planck cosmology parameters \citep{Planck2016}, these re-simulations follow the same cosmology. These cluster regions have been simulated with \RH{a range of simulation prescriptions including} \gadgetmusic\ \citep{Sembolini2013}, \gadgetx\ \citep{Rasia2015, Beck2016} and \simba\ \citep{Dave2019, Cui2022}. \RH{In this study, we will focus on the \simba\ result and leave a cross-model comparison for later work.}

The \simba\ run uses similar input physics to the {\sc Simba} simulation \citep{Dave2019}, which is an advanced version of {\sc Mufasa} \citep{Dave2016} based on the {\sc Gizmo} cosmological hydrodynamics code \citep{Hopkins2015} with its meshless finite mass hydro-solver. Because of the different simulation resolutions ($m_{\mathrm{gas}} \approx 2\times10^7$ \Msun\ for {\sc Simba} vs $\approx 2\times 10^8$~\Msun\ for \tth) and different objects (cosmological run for {\sc Simba} and galaxy clusters for \tth), the parameters for the baryon model are re-calibrated; see \cite{Cui2022} for detailed changes. Here, we only briefly list 
the baryon modelling included in \simba. Radiative cooling and photon-heating/ionization processes of gas are implemented using the Grackle-3.1 library \citep{Smith2017} with a spatially-uniform ultraviolet background model \citep{Haardt-Madau2012} and the self-shielding prescription following that in \cite{Rahmati2013}, and an H$_2$-based star formation model is included from {\sc Mufasa} \citep{Dave2016}. The decoupled two-phase model SN feedback is also adopted from {\sc Mufasa}, but with the mass loading factor from \citet{Angles-Alcazar2017b}, while the chemical enrichment model tracks eleven elements with metals from supernovae Type~Ia and Type~II and Asymptotic Giant Branch stars. More importantly, the AGN model adopts two BH accretion descriptions -- a torque-limited accretion model for the cold gas \citep{Alcazar2015, Alcazar2017a} and the Bondi accretion \citep{Bondi1952} for hot gas -- and three feedback modes: (1) `radiative mode' or `quasar mode' with bipolar outflows with velocities $\sim 500-1,500\kms$; (2) `jet mode' with bipolar ejection up to $15,000\kms$; and (3) X-ray feedback mode following \citet{Choi2012}. More details of the baryon models can be found in \citet{Dave2016, Dave2019}.

The AGN jet mode in the \textsc{Simba} model is particularly effective at ejecting and heating gas to large radii from halo centres, more so than in comparable cosmological hydrodynamic simulations \citep[for comparisons, see][]{Christiansen2020, Yang2022, Ayromlou2022, Cui2022}. It should be noted that the jet-mode feedback is implemented even more strongly in \tth\ run than the \textsc{Simba} cosmological volumes: jets are allowed to reach over twice-as-high velocities (maximally $15,000\kms$) than in \textsc{Simba} (maximally $7,000\kms$). This change was calibrated on the most massive clusters, necessary to achieve realistic quenching of cluster galaxies by $z=0$ \citep{Cui2022} -- however, the velocity formulae were applied identically in the lower-mass halos without a separate calibration. It is challenging to assess exactly how this affects the feedback with respect to the \textsc{Simba} cosmological volumes, as the jet speed varies rapidly but snapshots are only saved at coarsely-spaced time intervals. The overall picture is certainly that \tth\ \simba\ runs live on the extreme end of WHIM heating and gas ejection from lower-mass halos compared to other recent simulations like Illustris TNG. However, differences from other simulations does not signify inaccuracy. At the time of writing, there exists scant observational evidence for the gas properties of halos with $M<10^{14}$\Msun, especially at $z=1$, meaning the simulations assessed in this paper are viable models.

These simulations are analysed with two different algorithms to create halo catalogues: \caesar\footnote{\url{https://github.com/dnarayanan/caesar}} and The {\sc Amiga} Halo Finder\footnote{\url{http://popia.ft.uam.es/AHF/}} \citep[\ahf,][]{Gill2004, Knollmann2009}. The \caesar\ package identifies halos by a Friends-of-Friends (FOF) algorithm with linking length $b=0.2$. Such a linking length approximately leads to the identification of halos with an overdensity of 200 times the critical density \citep{White2001}, the same overdensity that is also frequently used to describe the virial radius \citep[e.g.,]{NFW1997}. The FOF halos are identified by firstly linking dark matter particles; then, all the other particles within the halo are identified by attaching each to the nearest dark matter particle. For \caesar\ FOF halos, the mass is given by the sum of all particles belonging to the halo. Galaxies in \caesar\ are identified through 6D FOF linking.

Meanwhile, \ahf\ defines halos based on the spherical overdensity method. Each halo has a position given by the halo's highest density peak, and a radius $R_{200c}$ within which the average density equals 200 times the critical density $\rho_c$. Thus, the halo mass is $M_{200c}$. Furthermore, the subhalos are identified through an unbinding process. However, we do not examine subhalos in this work. For larger halos hosting subhalos, the subhalo mass is included within the host halo mass.

As previously stated, we are mainly interested in results at $z=1$. The closest snapshot redshift is $z=0.987$; only simulation snapshots at that redshift are included in the main analysis and are referred to as $z=1$ for simplicity. A single brief comparison at $z=2$ is clearly marked.

\section{Methods}\label{sec:methods}

The oriented stacking method detailed in L22, which we will follow here, first creates many small cutouts from a Compton-$y$ map. Each cutout is centred on the location of a galaxy cluster. The large-scale structure surrounding the galaxy cluster location is then examined via an external galaxy catalogue. Gaussian smoothing is applied to the projected galaxies in a thin redshift bin surrounding the cluster redshift, and the smoothing scale is chosen based on the structures of interest. The curvature of the smoothed galaxy map provides the necessary information to determine the axis of strongest filamentary structure around each cluster (detailed in Sec.~\ref{subsec:orientation}). The $y$ map cutouts  are each individually rotated such that this filament axis is aligned across the entire sample. Finally, the sample of cutouts is stacked. The stacked image has a high SNR where structure is overlaid -- both in the centre, which displays the average galaxy cluster tSZ profile, and along the filament or supercluster axis, which averages over all contributions to the large-scale extended gas structure. The contributions include those from the intracluster medium in neighboring clusters, the intragroup medium in intermediate-mass halos, the circumgalactic medium around individual galaxies, and the WHIM.
\subsection{Halo selection}\label{sec:halo_sel}

To create a similar setup to previous and ongoing work with observational data (L22), we choose to examine the anisotropy of structure around massive halos which would likely be identified as galaxy clusters in optical or SZ surveys \citep[e.g.,][]{Hilton2021, Rozo2014}. We select snapshots that contain halos at $z=1$ with $M>10^{14}$\hMsun; although clusters with lower masses can be identified with optical surveys, such low-richness cluster samples are typically more contaminated by projection effects \citep{McClintock2019}. Of the 324 simulations, most (301) have formed such massive halos by $z=1$, satisfying this criterion.

A further criterion is applied based on how close the massive halos are to the box centre at $z=1$. The simulations are set up such that the most massive halo is guaranteed to be at the centre by $z=0$, but at $z=1$ its progenitor(s) are often still far from the centre. We wish to examine only those simulations for which a massive halo is already near the centre by $z=1$, so that when examining the $z=1$ snapshots, we can analyze the mock tSZ signal out to a large extent around each halo without concerns about projection and contamination effects in the outer regions of the simulations (further discussed in Sec.~\ref{subsec:y-map}). To do so, we further limit the sample to only those snapshots in which an $M>10^{14}$~\hMsun\ halo is within 3~\hMpc\ from the centre. Many more snapshots are removed from this second cut; the final sample consists of 98. The central halo masses range across snapshots from $1.1-6.7\times10^{14}$~\hMsun\ with a distribution skewed towards lower masses. Because the second position-related cut dominates the selection, and is arbitrary with respect to halo properties, the mass function of all halos across all snapshots in the selected sample is nearly identical to that of the full sample. Furthermore, we check for bias in several environmental properties: the smoothed 2D overdensity and ellipticity of the galaxy field at a scale of 10~\hMpc\ (with the ellipticity measured as a difference between eigenvectors of the Hessian matrix at the snapshot centre). We find no sign of significant differences in the selected sample, confirming that the selection does not induce bias in our results.


\subsection{Halo gas vs diffuse gas}\label{sec:halo_diffuse}

\begin{figure*}
    \centering
    \subfloat{\includegraphics[width=0.49\textwidth ,trim={0cm .25cm 0cm 0cm},clip]{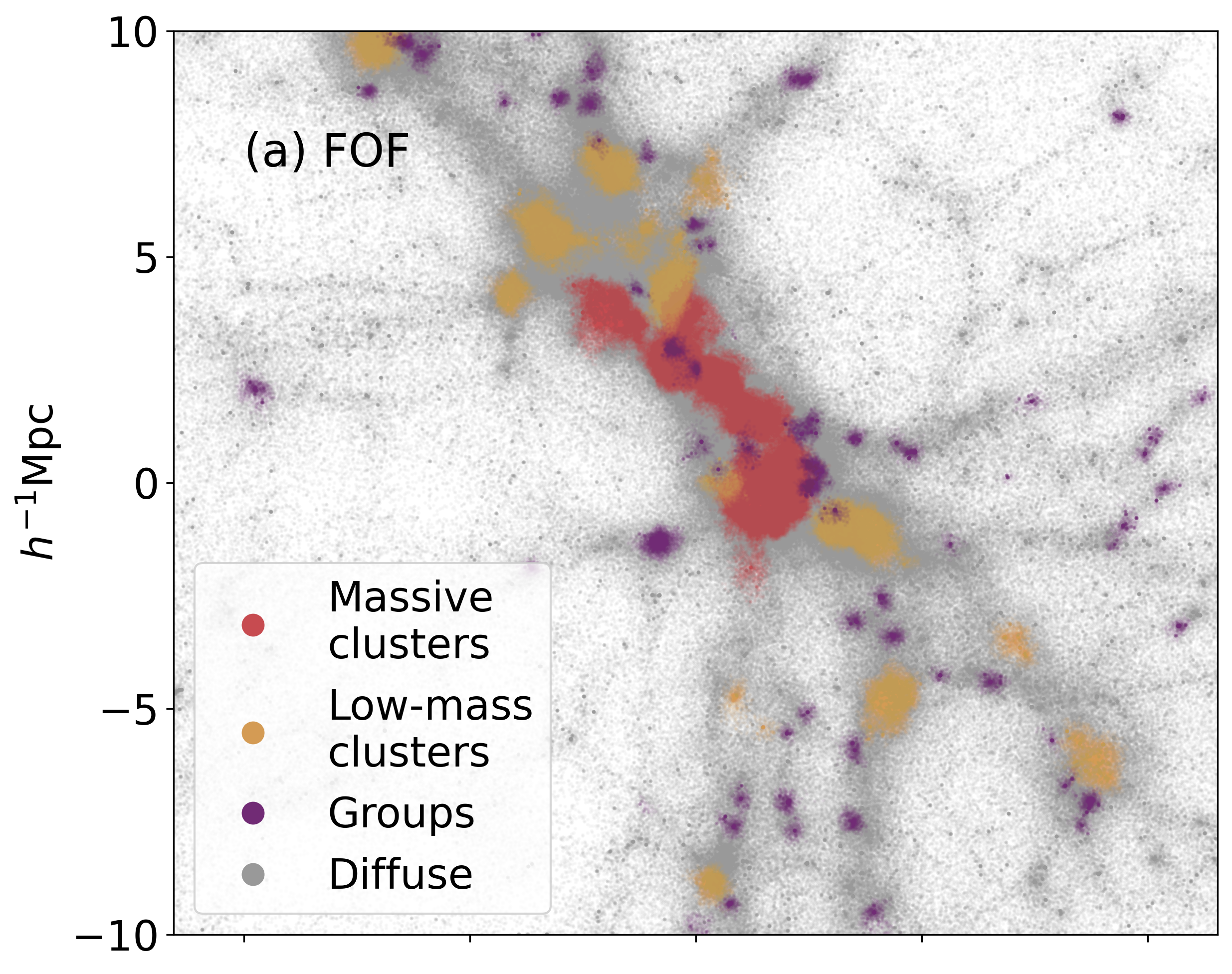}}\hspace{1em}
    \subfloat{\includegraphics[width=0.43\textwidth]{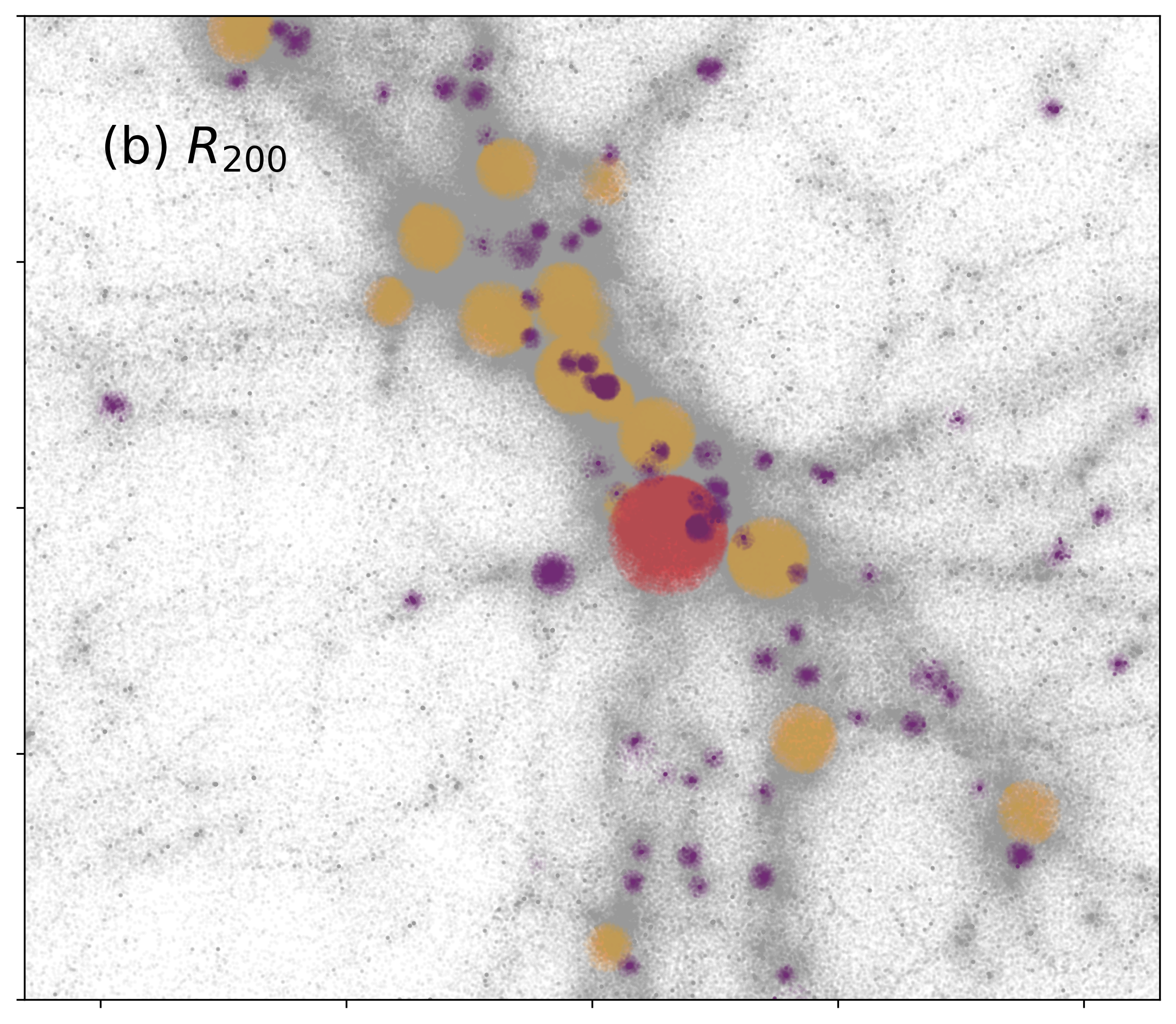}}\\
    \subfloat{\includegraphics[width=0.49\textwidth]{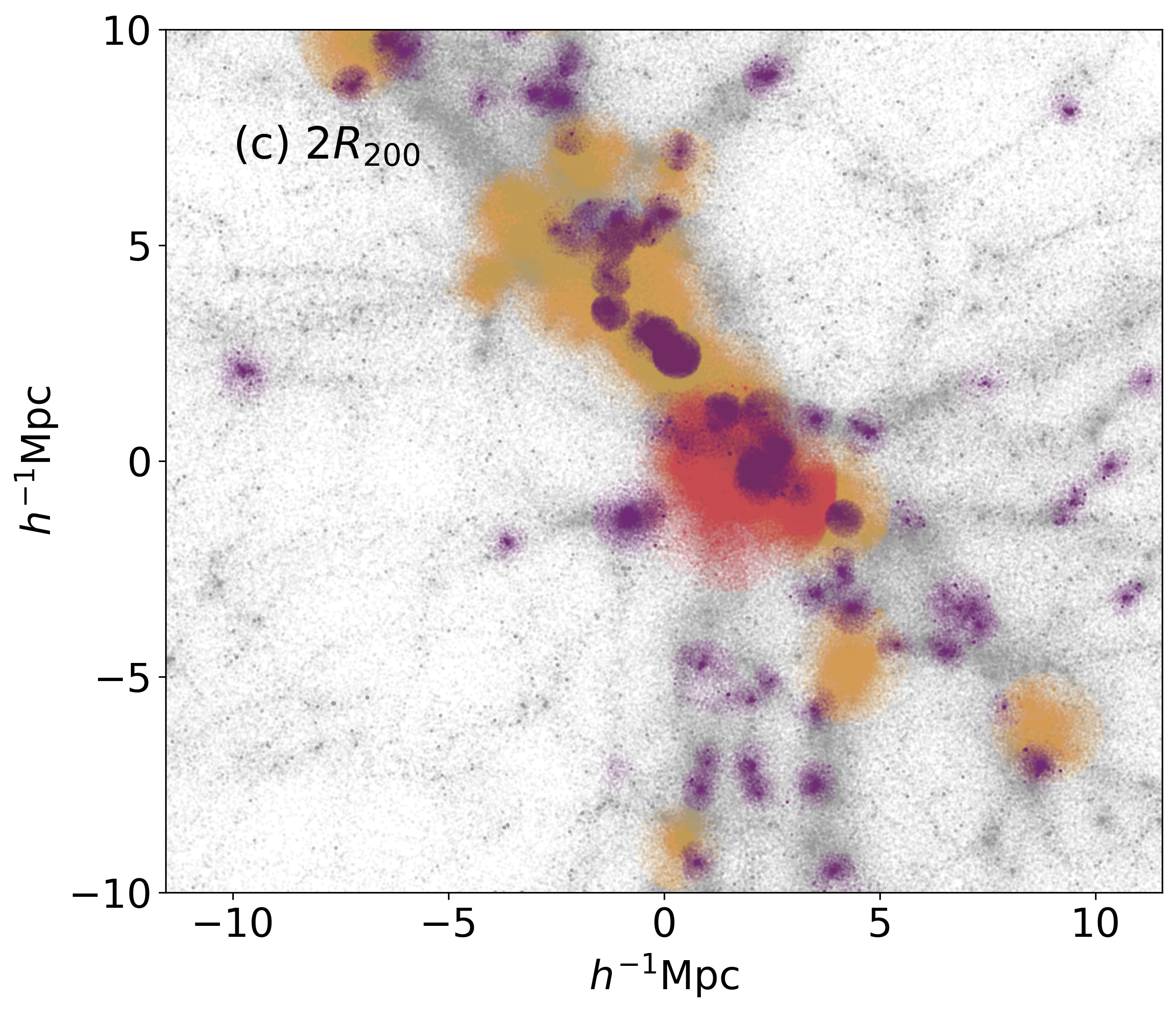}}\hspace{1em}
    \subfloat{\includegraphics[width=0.43\textwidth]{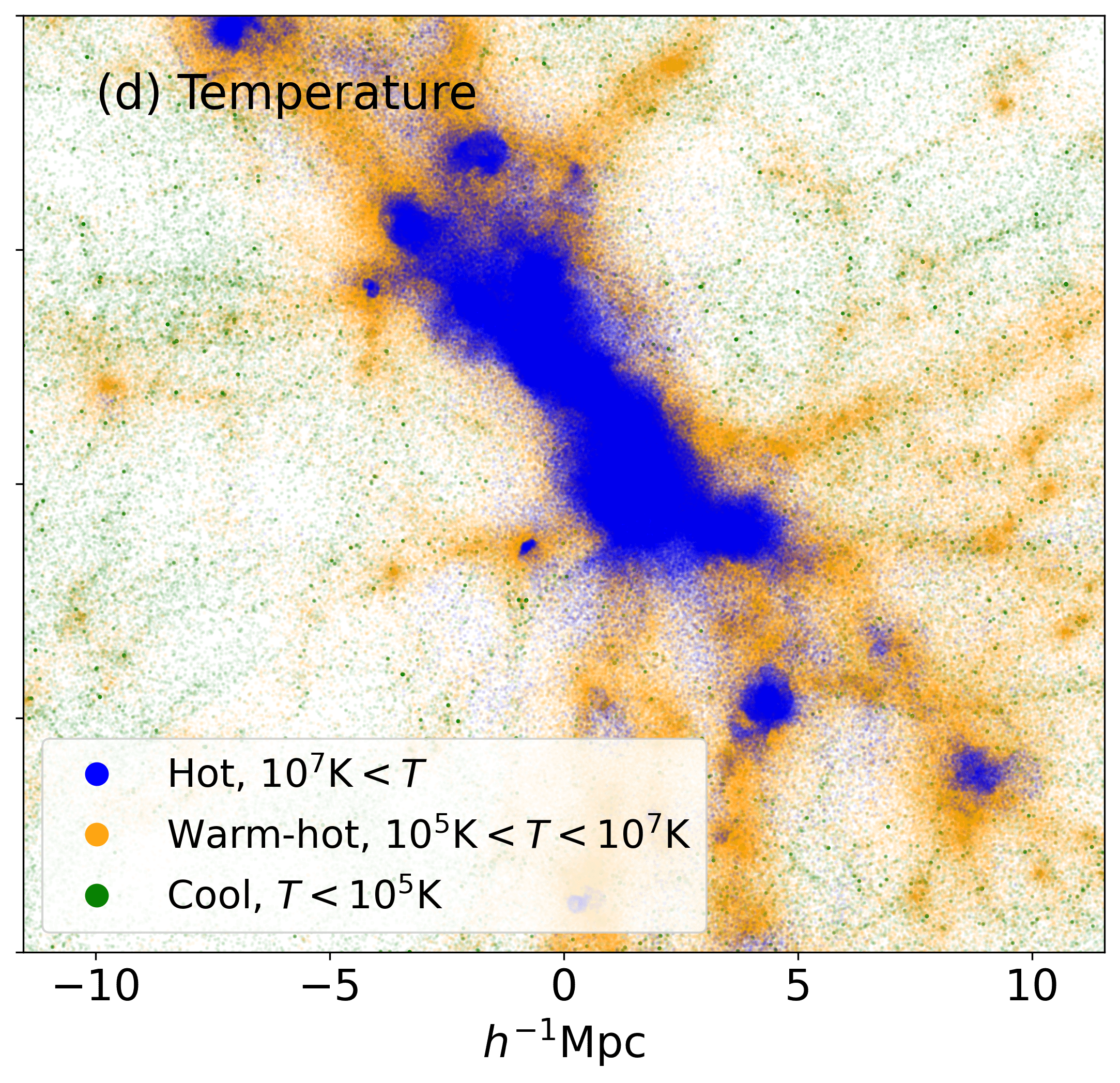}}
    \caption{Comparison of the gas particles identified as associated with halos of different mass ranges by three separate methods. All plots show a slice 14~\hMpc\ deep in the $z$ direction of the same snapshot, with 75\% of particles plotted for each category. Starting from the upper left, gas particles are considered halo-associated if (a): they are identified by friends-of-friends as halo members (these particles make up 11\% of the total gas particles); (b): they lie within $R_\mathrm{200c}$ of halos (12\% of particles); (c): they lie within $2 R_\mathrm{200c}$ (25\% of particles). In (d), particles are divided by temperature.}
    \label{fig:bound_unbound}
\end{figure*}

The WHIM is generally defined as gas in the temperature range $10^5<T<10^7$~K \citep{Dave2001} which mostly exists in a non-virialized state beyond halos. The baryons can be outside of halos either because they have not yet accreted onto halos or because they were blown out by feedback processes.

A particular question of interest is whether simulations which paste pressure profiles onto halos can adequately capture the contribution of the diffuse gas to the tSZ. This technique is easily applied in post-processing to N-body simulations \citep[e.g., Buzzard,][]{DeRose2019} or rapid predictive simulations \citep[e.g., Websky,][]{Stein2019} but makes two assumptions \RH{that are not strictly true}: 1) that the profiles are isotropic and (2) that the gas pressure drops sharply to zero at some radius. The second assumption is necessary to avoid excessive overlap between pressure profiles from neighboring clusters. A commonly used pressure profile model was fit to clusters from hydrodynamic simulations with AGN feedback from \citep{Battaglia2012a,Battaglia2012b}, which simulated clusters down to a minimum halo mass of 1.4$\times10^{13}$\Msun. In recent applications (e.g., L22) using the Websky and Buzzard simulations, this profile model was applied to halos down to $M\sim10^{12}$\Msun\ with an extent of $4\times R_{200c}$. However, it is unclear whether the model is trustworthy at this radius or for this large mass range, and the highly extended profile almost certainly causes double-counting of $y$ signal from nearby halos \citep{Stein2020, Battaglia2012b}.

Therefore, several key questions are: \textit{how much tSZ signal is contributed by halos and diffuse gas below 1.4$\times10^{13}$\Msun? How much tSZ signal is contributed by the outskirts of halos, and how far do such profiles need to extend to capture sufficient $y$ signal without over-counting?} \RH{The larger volume of \tth\ compared to \citet{Battaglia2012a} and the improved mass resolution allows for the investigation of these questions directly.}

We begin by dividing the gas particles in each snapshot into different categories. First, star-forming particles and wind particles\footnote{The particle is recently ejected by feedback and currently detached from hydrodynamical calculation.} are removed. Next, due to the ambiguity in dividing concentrated and diffuse gas, we divide the gas into regimes with several different approaches. The first approach splits the gas by a simple temperature cut. We label gas with $10^5 <T<10^7$~K as `warm-hot', following the original definition of the WHIM gas by \citet{CenOstriker1999, Dave2001}. We exclusively refer to this as `warm-hot gas' rather than the `WHIM' to avoid confusion with other studies which sometimes apply a density cut in addition to the temperature cut. The warm-hot category includes gas both within and beyond halos, but $\sim90\%$ of its mass is beyond $R_{200c}$ of halos making it mostly diffuse. We label gas with $T>10^5$~K as `hot', over half of which is typically within $R_{200c}$ of halos. These categories are shown for one snapshot in Fig.~\ref{fig:bound_unbound} (d). In addition, we show the distribution of cool gas with $T<10^5$~K for visual purposes, but elect not to report results for it in this paper as we find it contributes negligibly to the tSZ effect.

The other approach divides the gas into particles associated with halos and those beyond halos. We divide the halos into three mass ranges outlined in Table~\ref{tab:halo_mass_ranges}, corresponding roughly to massive clusters, low-mass clusters, and groups. Halos with lower masses, $M<10^{12}$\hMsun, are poorly resolved in the simulation (sampled by fewer than 1000 particles), and thus we combine those particles along with the diffuse gas which is unassociated to any halo into a fourth category. This is also a useful division observationally, as $M\sim10^{12} h^{-1}$\Msun\ is approximately the lowest mass probed by any currently available large galaxy survey \citep[e.g., CMASS, as shown in][]{Schaan2021}.

\begin{table}
    \centering
    \begin{tabular}{c|c}
    Label & Gas definition\\
    \hline
    Warm-hot & $10^5<T<10^7$~K \\
    Hot & $10^7~\mathrm{K}<T$ \\
    Massive clusters & Assoc. with $M>10^{14} h^{-1}$ \Msun\, halos \\
    Low-mass clusters & Assoc. with $10^{13}<M<10^{14} h^{-1}$ \Msun\, halos \\
    Groups & Assoc. with $10^{12}<M<10^{13} h^{-1}$\Msun\, halos\\
    Diffuse & Unassoc. with any $M>10^{12}$~\hMsun\ halo
    \\
    \end{tabular}
    \caption{Definitions used to define different gas particle categories. `Associated' (assoc.) refers to different definitions as detailed in Table~\ref{tab:bound_def}.}
    \label{tab:halo_mass_ranges}
\end{table}

\begin{table*}
    \centering
    \begin{tabular}{c|c|c}
    Label & Definition & Diffuse gas mass fraction\\
    \hline
    FOF & Gas particles identified by FOF as halo members & 0.87\\
    $R_{200c}$ & Particles with $|\boldsymbol{x}-\boldsymbol{x}(\Phi_{min})|$< $R_{200c}$&0.87\\
    $2 R_{200c}$ & Particles with $|\boldsymbol{x}-\boldsymbol{x}(\Phi_{min})|$< $2 R_{200c}$&0.72\\
    \end{tabular}
    \caption{Variations in the definition of halo-associated gas. $\Phi_{min}$ refers to the minimum potential of the halo and $\mathbf{x}$ to the 3D position coordinates of each particle. For each definition, the third column shows the mass fraction of the gas beyond halos, which includes all gas particles not encompassed by the halo-associated definition. The majority of the gas mass is beyond halos. For comparison, when divided by temperature alone, warm-hot gas (both within and beyond halos) constitutes 60\% of the total mass.}
    \label{tab:bound_def}
\end{table*}

To define particles that are associated with the halos in each mass range, we use multiple approaches as listed in Table~\ref{tab:bound_def}. The first approach simply selects those which are identified as belonging to halos of that mass range by FOF in \caesar. The second defines halo particles as those which lie within $R_{200c}$ from the centre-of-mass of halos defined via \ahf. Such gas is sometimes loosely referred to as `bound' in the literature, as $R_{200c}$ is approximately the halo virial radius, although the true bound/unbound nature depends on the gas velocity, which is more complex than that of the dark matter due to feedback. Additionally, to probe further out from each halo, we use a wider limit of $2 R_{200c}$ for a third comparison. We consider extending the threshold further (e.g., $4R_{200c}$), but the limited snapshot size makes this unfeasible, as too few particles in the central uncontaminated region remain. Table~\ref{tab:bound_def} also lists the fraction of gas mass in the diffuse state under each definition; 87\% of the gas lies beyond $R_{200c}$ of halos, which is consistent with the findings of \citet{Sorini2022} for the \textsc{Simba} simulations.

Figure \ref{fig:bound_unbound} shows the gas particle divisions via different methods for one snapshot. There is some overlap between particles which lie within $R_{200c}$ and especially $2 R_{200c}$ of halo centres which belong to different mass ranges, therefore the mass ranges in Table \ref{tab:halo_mass_ranges} are not entirely distinct. In all cases, the diffuse particle category \textit{is entirely distinct}, as it contains all particles which are not considered associated with any halo above $10^{12}$\hMsun. The different methodology has a clear visual impact on how the gas from different categories is distributed; the gas associated with high-mass FOF halos in this snapshot, for example, is far more extended than \RH{the gas} associated to high-mass halos by the virial radius. This suggests that the assumed definition is important when discussing the contributions of diffuse gas \RH{to the tSZ signal}.

\begin{figure}
    \centering
    \includegraphics[width=\columnwidth]{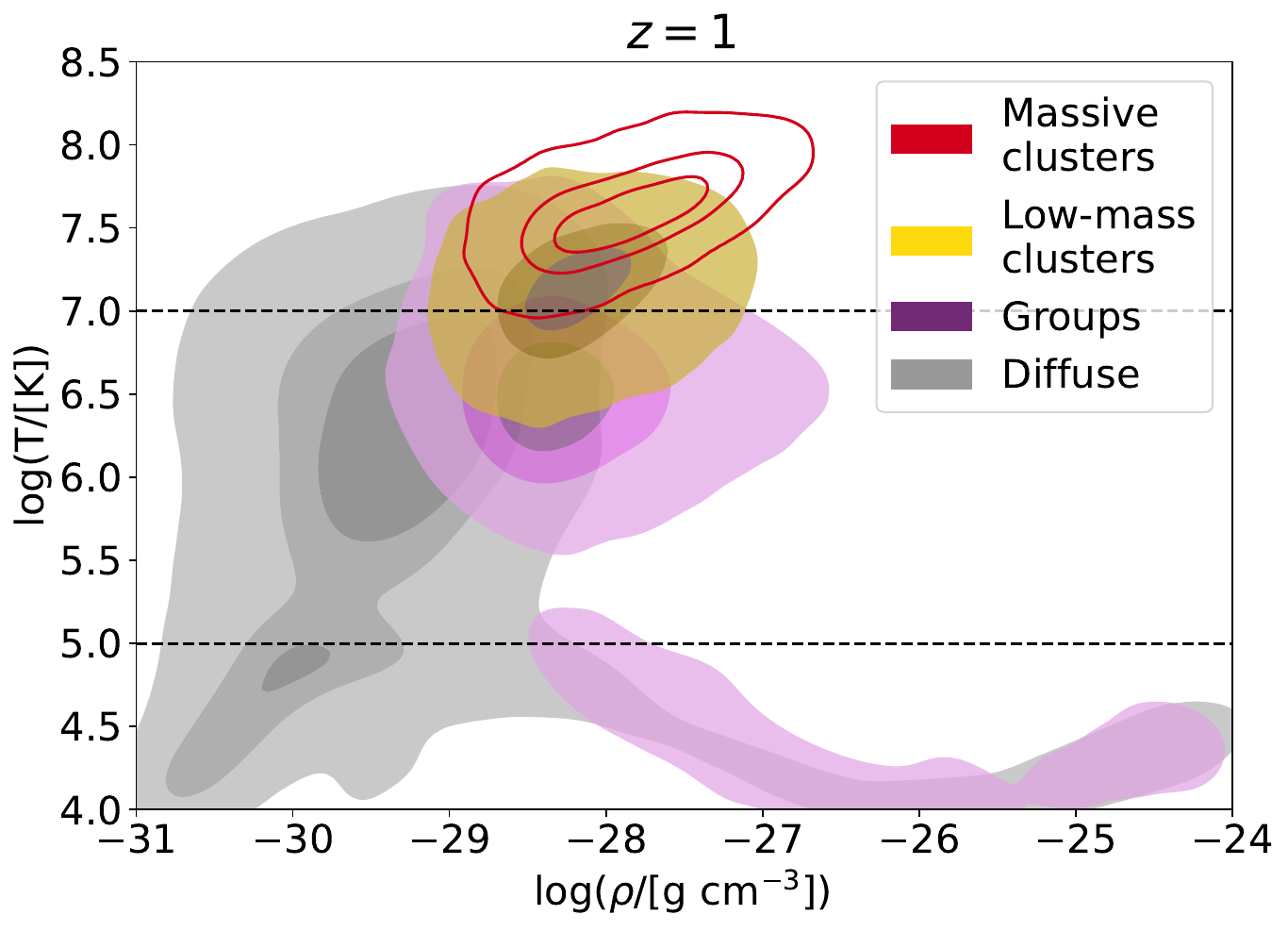}
    \includegraphics[width=\columnwidth]{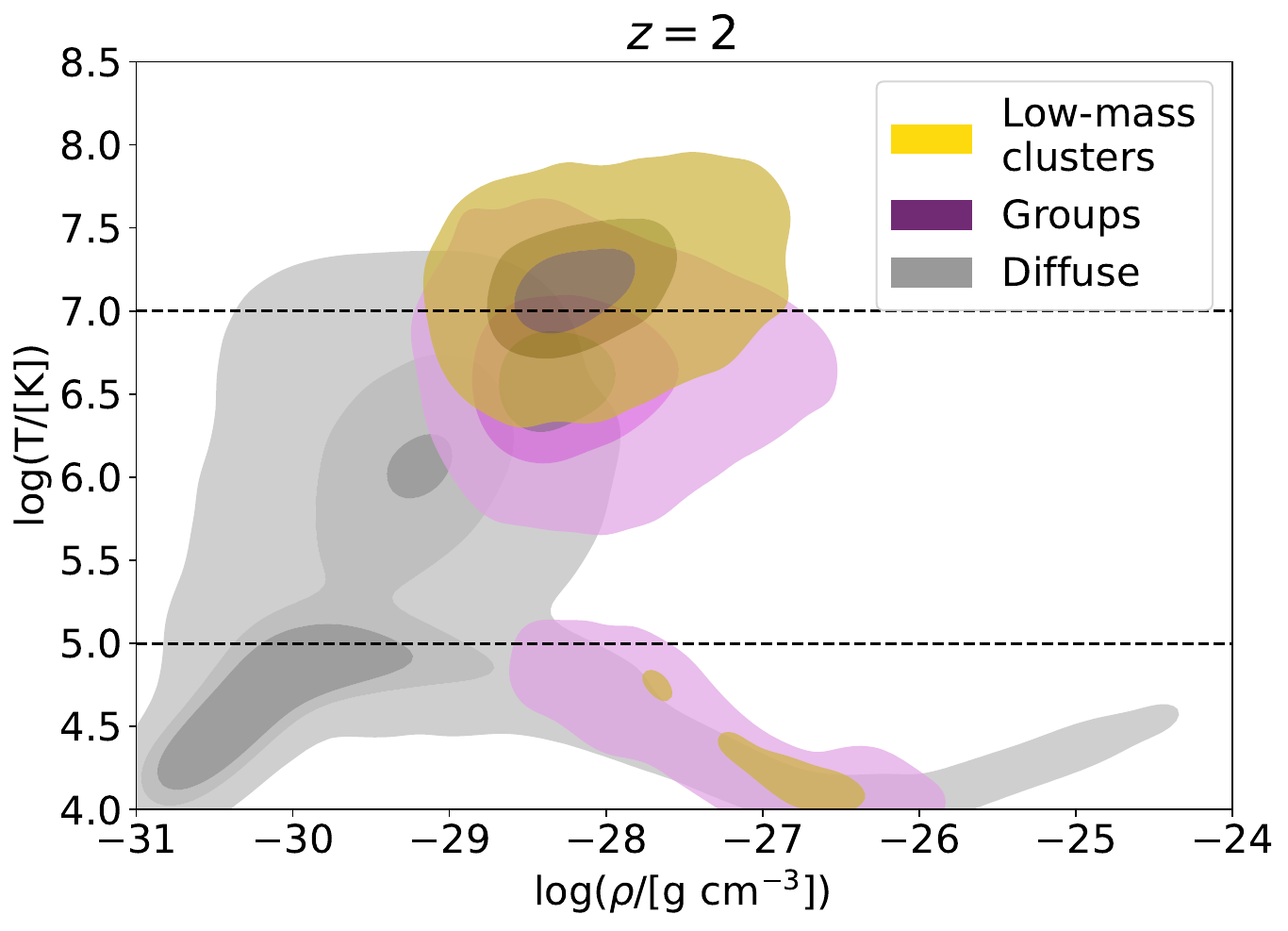}
    \caption{Comparison of the temperature and density of gas particles identified as associated with halos of different mass ranges by the $R_{200c}$ method at $z=1$ (top) and $z=2$ (bottom). Each distribution represents a random $0.1\%$ of all particles in the given category drawn from all 98 chosen snapshots. Dashed black lines indicate the lower and upper bounds in temperature for the `warm-hot' gas. The high-density trail at $10^4$ K corresponds to star formation. The biggest differences as time evolves from $z=2$ to $z=1$ are the creation of massive clusters and the heating of diffuse gas to the warm-hot phase.}
    \label{fig:gas_phases}
\end{figure}

The distribution of temperature and density for the particles identified via $R_{200c}$ is shown in the upper panel of Fig.~\ref{fig:gas_phases}. Although categories overlap, the combined temperature and density (the pressure) is highest in massive galaxy clusters, followed by low-mass clusters, groups, and the poorly-resolved plus diffuse gas. The high-density, low-temperature track corresponds to star formation: although most star-forming particles are cut, the \RH{particles that remain} are likely those which have just begun star-forming but have not yet updated their star formation tag in the simulation. Despite the high densities, we confirm that the low temperature of these particles results in a negligible contribution to $y$ compared to signals from any other category in this work. Thus, inclusion of these particles in any of the halo-associated categories does not bias the resulting signals, and we elect not to present any separate results for these `cool gas' particles.

For the division using $2 R_{200c}$ (not shown in Figure~\ref{fig:gas_phases}), the distributions of temperature and density are similar but there is more overlap between the halo categories, as the halo outskirts encompass lower-pressure gas. Overall, the halo gas extends to lower pressures, and the smaller amount of remaining diffuse gas is thus also concentrated at slightly lower densities and temperatures than in the $R_{200c}$ case. For the FOF case, there is slightly less overlap overall than either of the previous cases. In all cases, the warm-hot temperature range encompasses mostly gas from the `diffuse' and  `group' categories.

For a point of comparison that indicates how these distributions evolve over cosmic time, we also show the same plot for the gas phase at $z=2$ in the lower panel of Fig.~\ref{fig:gas_phases}. Compared with $z=1$, a notable difference is that there are no $M>10^{14}$\hMsun\ halos at this redshift; their predecessors are the low-mass clusters. The diffuse and unresolved gas at this redshift is shifted towards lower temperatures and densities; most of it is not yet in the warm-hot phase by this point in time. The warm-hot regime at $z=2$, then, is dominated by groups and the tail-end of low-mass clusters. AGN feedback and shock-heating of gas both contribute to the shifting of the diffuse gas from $z=2$ to $z=1$ \citep{Sorini2022}.

\begin{figure*}
    \centering
    \includegraphics[width=1.6\columnwidth]{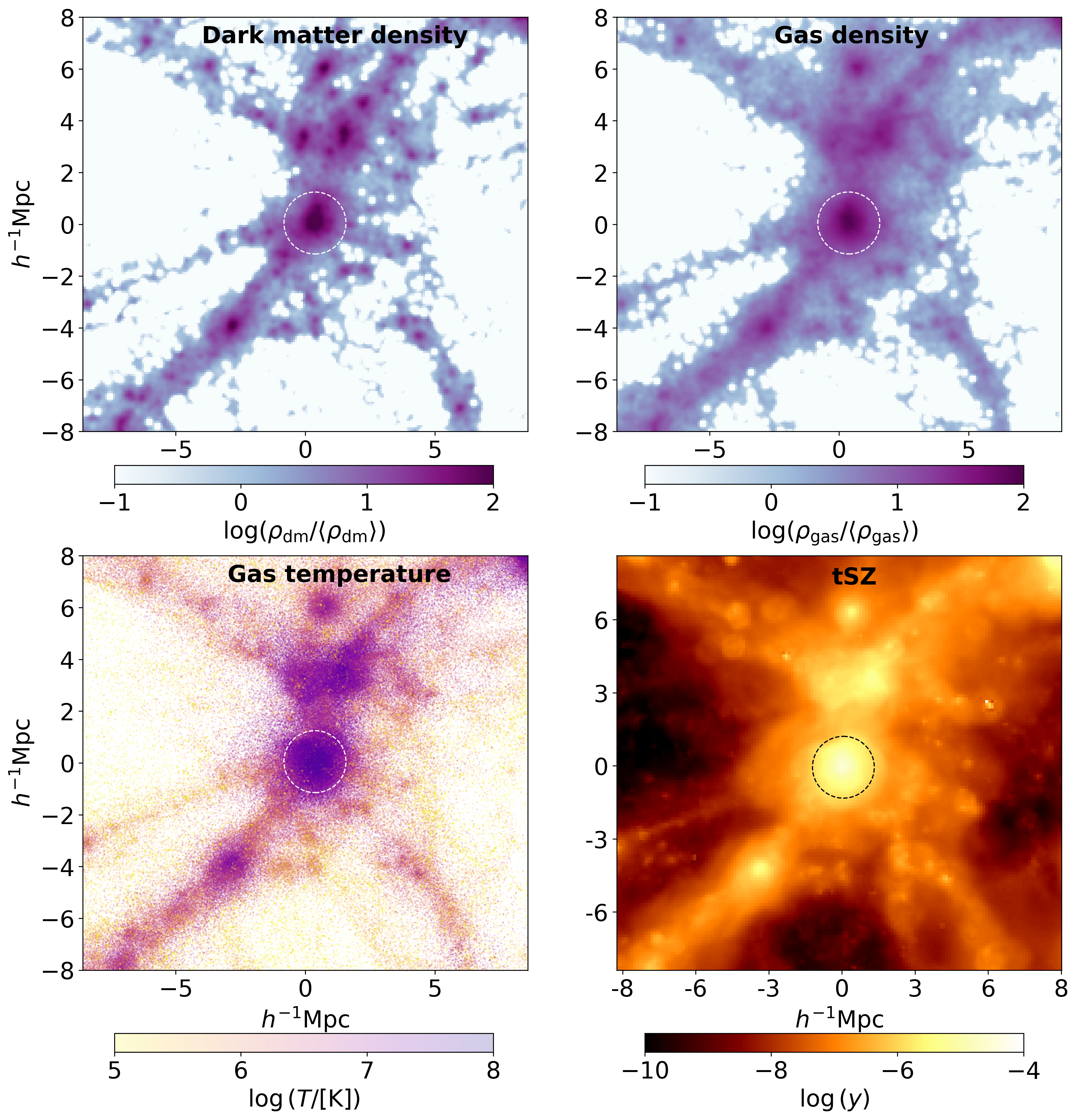}
    \caption{Clockwise from upper left: the 2D dark matter density contrast, calculated in projection over a 14~\hMpc\ slice; the same for gas density; Compton-$y$ map; and gas temperature. The dashed circles represent $R_{200c}$ of the central halo. The density contrast colour bar is limited to allow for visual comparison of low-density regions; dark purple regions are highly over-saturated. The dark matter halos are more concentrated and anisotropic than the gas, which has a puffy sphericalized profile around halo locations extending beyond $R_{200c}$ in both the density and gas temperature, and consequentially in tSZ.}
    \label{fig:ymap_creation}
\end{figure*}

The density-temperature distributions provide some indication of how each component will contribute to the observed tSZ signal, because the pressure determines the $y$ contributed per unit length along the line-of-sight. However, as the physical extent of the gas also affects the resulting observed tSZ, the phase distribution does not tell the entire story.

\subsection{y map creation} \label{subsec:y-map}

With each subset of particles, we create $y$ maps using the \texttt{pyMSZ} package\footnote{\url{https://github.com/weiguangcui/pymsz}}. More details of this package can be found in \citet{Baldi2018,Cui2018}. The calculation basically follows Equation~\ref{eq:tSZ}; the integration is represented by the summation of all the gas particles in simulations. To be consistent, we also adopt the same \RH{smoothing kernel as the smoothed particle hydrodynamics (SPH) simulations} to smear the $y$ signal from each gas particle to the projected image pixels. Note that, due to the unrealistic model treatments, we exclude the star-forming gas and wind particles.

\begin{figure*}
    \centering
    \subfloat[]{\includegraphics[width=0.3\textwidth, trim={.5cm 0.5cm 0cm 0.5cm},clip]{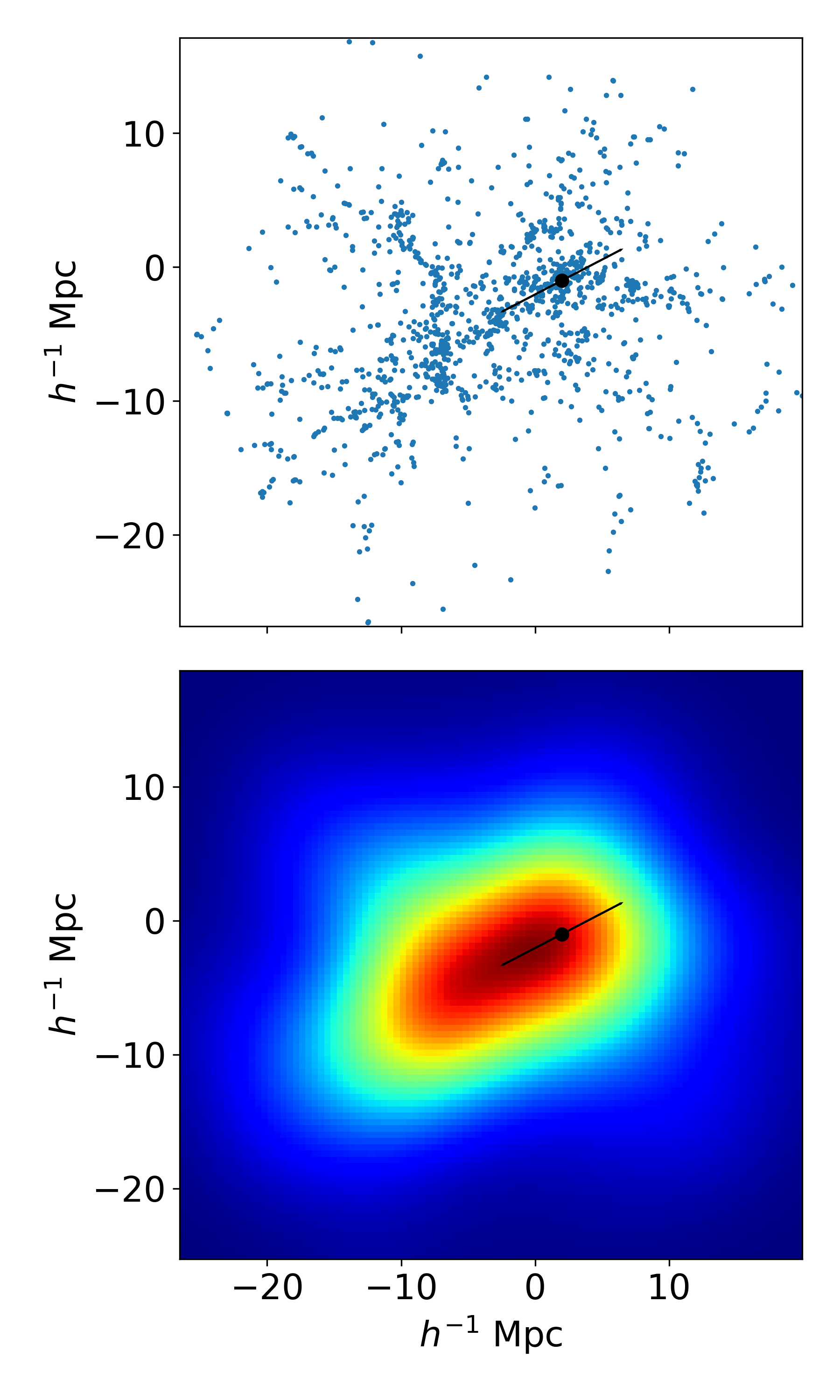}}
    \subfloat[]{\includegraphics[width=0.3\textwidth, trim={.5cm 0.5cm 0cm 0.5cm},clip]{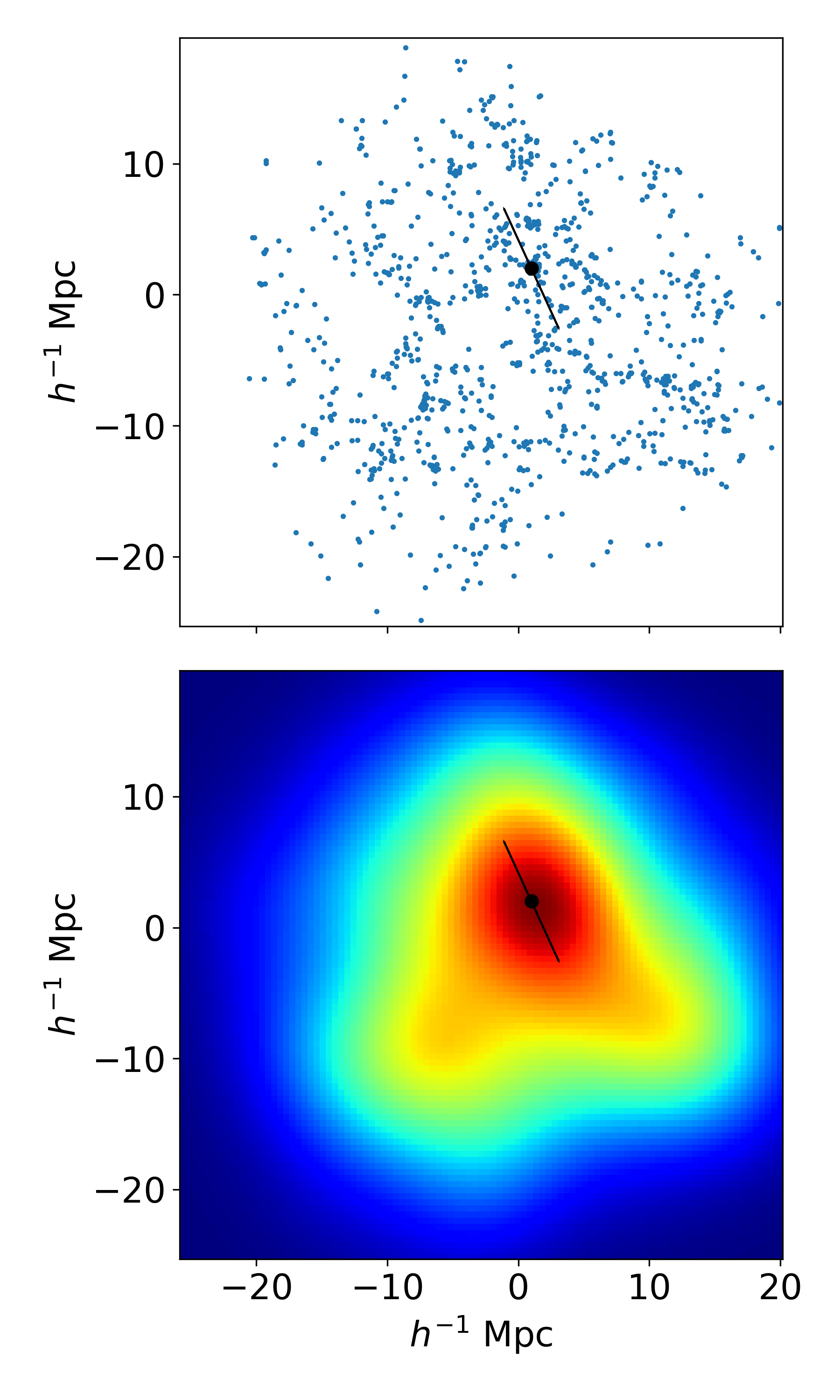}}
    \subfloat[]{\includegraphics[width=0.3\textwidth, trim={.5cm 0.5cm 0cm 0.5cm},clip]{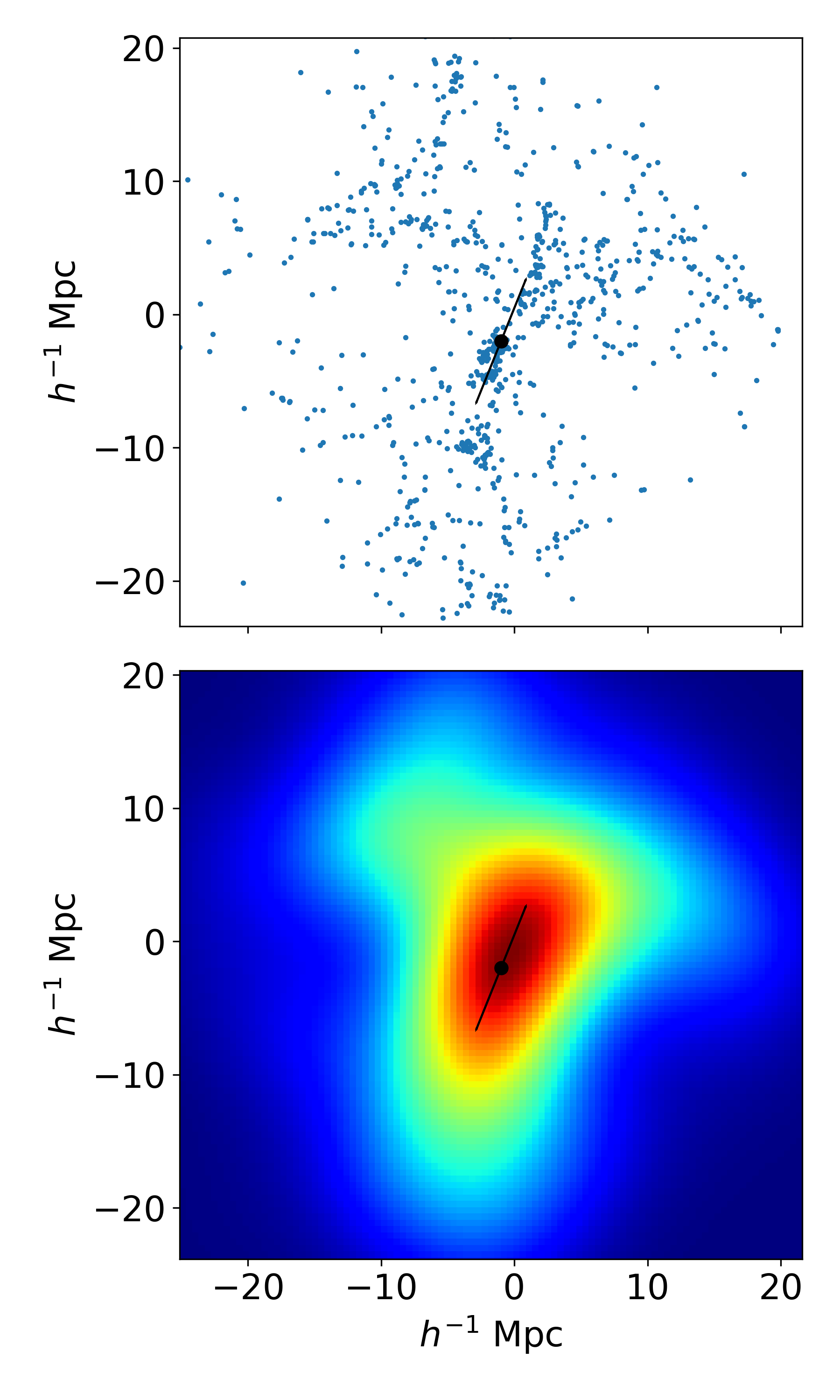}}
    \caption{Galaxies (top) and the smoothed galaxy field $\tilde{n_g}$ (bottom) for three snapshots. Black points show the positions of the massive clusters that will be at the centre of each respective $y$ map; the black lines show the direction of $\boldsymbol{v_2}$ with a length equal to the FWHM of smoothing.}
    \label{fig:orientation}
\end{figure*}

For each snapshot, we create a Compton-$y$ map projected along the $z$ direction. We centre the $y$ map on the position of the selected massive halo. We check for contamination from the low-resolution dark matter particles beyond the inner high-resolution region by examining the ratio of gas particle number to low-resolution dark matter particle number in each cell. Cells containing more than 0.1\% low-resolution particles are considered contaminated. At $z=1$, the percentage of contaminated cells is typically $<1\%$ out to 20~\hMpc, and increases closer to the snapshot edges. Thus, we choose to include data only within a radius of 20~\hMpc. The $y$ map is created by integrating Eq.~\ref{eq:tSZ} along the line-of-site across a chosen extent. Because the snapshot data are distributed somewhat spherically, integrating $y$ throughout the entire snapshot would cause an artificial decrease in $y$ from the centre to the outskirts due to the decreasing line-of-sight extent of the sphere. We thus limit $d\ell$ to range from [-10,10]~\hMpc\ along the $z$ axis. To ensure that all parts of the rectangular box with this depth fit within the uncontaminated sphere of 20~\hMpc\ radius, and also taking into account the small shifts from centreing on the massive halo, we only examine results out to 8~\hMpc\ in radius from the centre of the resulting images. 

We create the $y$ maps at 10 arcsecond resolution to produce detailed images for visualization purposes; later, for image analysis we degrade the resolution to $\sim40$ arcsecond for faster computation.

Fig.~\ref{fig:ymap_creation} demonstrates the making of a $y$ map. The dark matter density is shown in the upper left to demonstrate how it is more concentrated and asymmetric than the gas density (upper right). The gas density and temperature (lower left), also quite isotropically distributed around halos, both enter into the $y$ map (lower right).

\subsection{Orientation}\label{subsec:orientation}

Observationally, Compton-$y$ maps are noise-dominated and thus the filament axis cannot be determined by the $y$ map itself; another large scale structure tracer must be used. Maps of galaxies from large surveys are currently the best tracers for filamentary structure. In simulations, more accurate orientation (with respect to the true density field) can be achieved by weighting galaxies by their mass, or even incorporating the dark matter particles from the snapshot. However, as individual galaxy masses are often poorly constrained or unreported in large survey data, and the best observational proxy for the dark matter (lensing) is noisy, in this work we use the galaxy number density to create a setup that is most straightforwardly performed on observational data. Additionally, although full 3D information is accessible with simulations, the currently available galaxy data with sufficiently high number density for oriented stacking comes from photometric surveys \citep[e.g. DES Y3,][]{SevillaNoarbe2021, Porredon2021}. Given photometric redshift uncertainties, \RH{$\sigma_z \sim 0.01(1+z)$,} the best approach is to project galaxies into tomographic bins of width $\sim100-200$~Mpc.

Following the observational motivations, we use the simulated \caesar\ galaxy data to map galaxies in projection along the $z$-axis (Fig.~\ref{fig:orientation}, top). We then smooth the maps with a Gaussian filter with a chosen scale discussed below (Fig.~\ref{fig:orientation}, bottom). Despite mirroring the observational approach, several differences arise when using these simulations. Due to the limited snapshot size, there is limited projection from uncorrelated structure compared to, e.g., the 200 Mpc bins used in L22 (the 30~\hMpc\ extent is about 1/3 of the bin width). Orientation with the full \caesar\, galaxy sample is also more accurate than with a real photometric galaxy sample, as we do not attempt to introduce any contamination to the sample to mimic the effects of photometric redshift scatter. Furthermore, the orientation will be more accurate than with extant spectroscopic samples, as current best large-survey galaxy samples are magnitude-limited \citep[e.g., CMASS galaxies are limited to $M^*\sim10^{10.5}$~\Msun\ or higher,][]{Maraston2013}, while 
we include all \caesar\ galaxies (with stellar masses as small as $\sim10^{10}$~\Msun).

\begin{figure*}
    \centering
    \includegraphics[width=1.8\columnwidth, trim={0cm 1cm 0cm 0.5cm},clip]{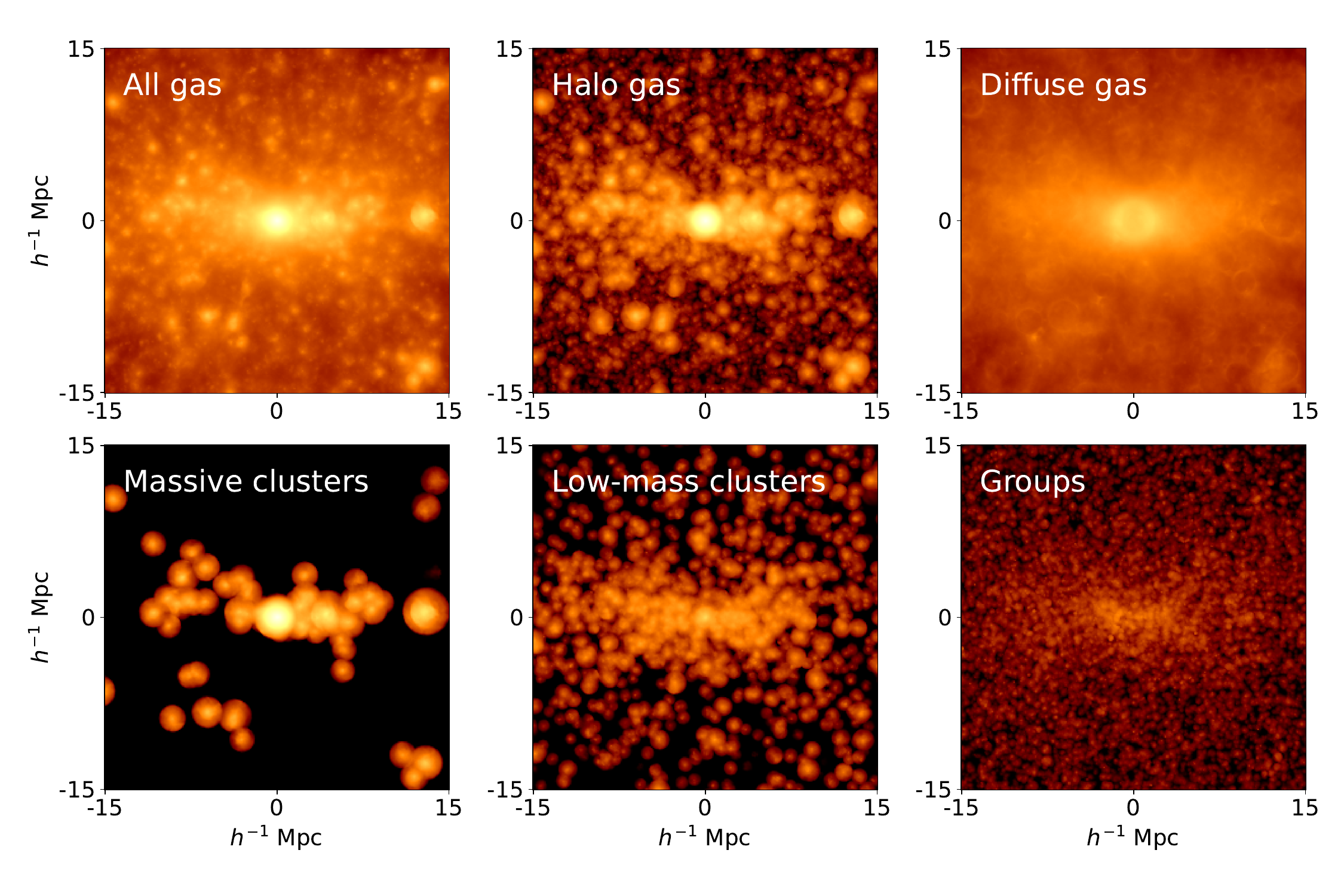}
    \includegraphics[width=1.2\columnwidth]{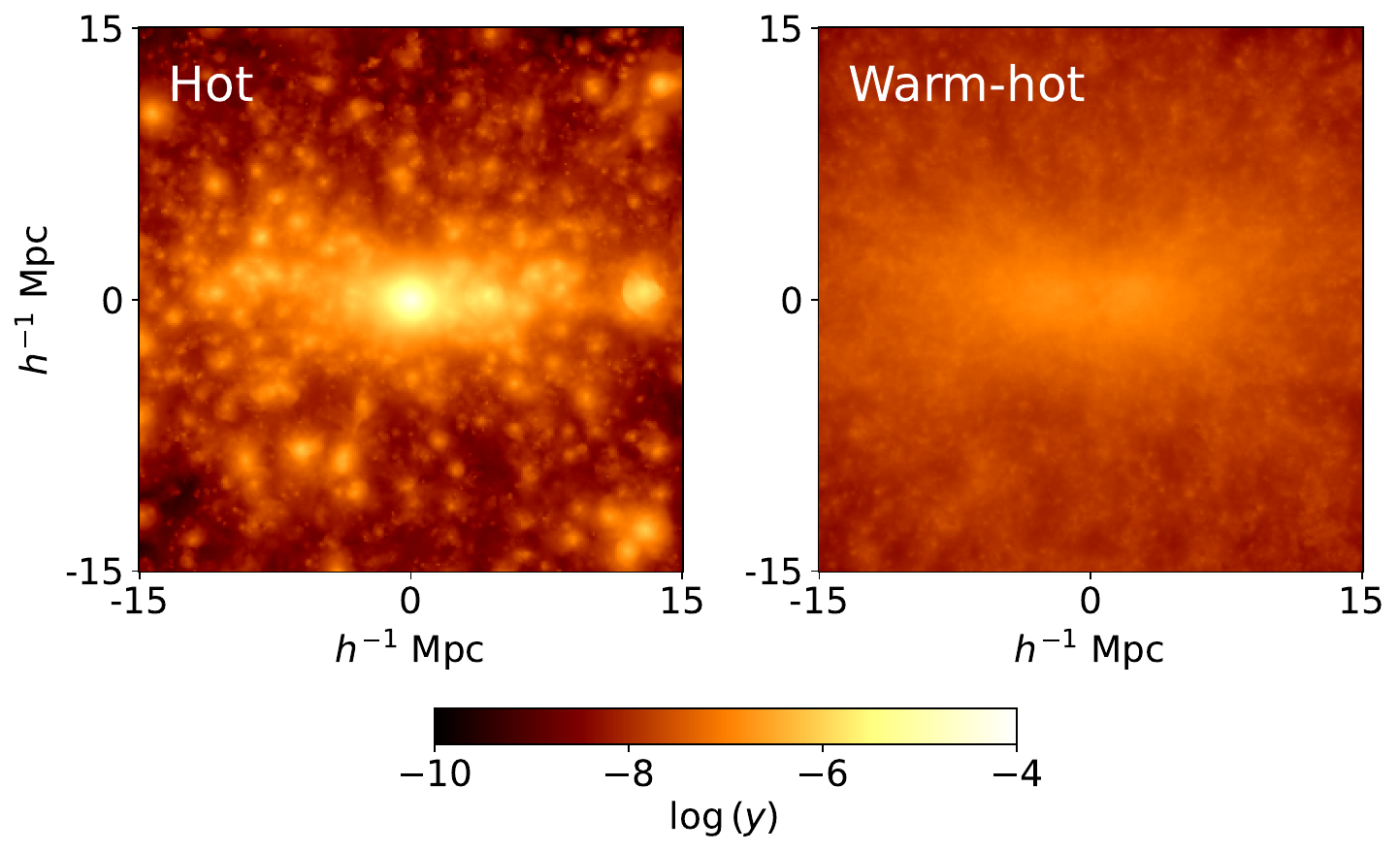}
    \caption{$y$ map stacks, each combining oriented $y$ maps from of 98 separate snapshots, including different particle subsets. Top: all gas particles (left), the gas within $R_{200c}$ of halos (middle), and the gas beyond $R_{200c}$ (right). The halo and diffuse particles are mutually exclusive sets, so the top-middle and top-right maps sum to make the top-left map. Middle: The halo gas particles corresponding to different mass ranges. There is some physical overlap between halos in the different categories such that some gas particles are repeated in multiple mass categories. When summed, the middle plots make a $y$ map with slightly more signal than the `all halo gas' map. Bottom: hot gas ($T>10^7$K) and warm-hot gas ($10^5$K$<T<10^7$K). We elect not to show the cool gas ($T<10^5$K) as its contribution is so small it is not visible with this colour bar. The diffuse and warm-hot gas appear less concentrated around the filament axis than the halo gas / hot particles. Note that the colour bar is in log scale in order to show the weaker $y$ signals from groups and diffuse gas; these signals are orders of magnitude below the $y$ in clusters.}
    \label{fig:allstacks_r200}
\end{figure*}

The scales explored in this work are motivated as follows. Typical filaments bridging two clusters project into a transverse comoving distance range of $\sim6-14 h^{-1}$  Mpc \citep{degraaff2019}. When centreing an analysis of orientation on a cluster rather than examining cluster-pairs, it is therefore logical to examine a smoothed field that probes at least 6$h^{-1}$ Mpc in any direction from the central cluster. The \tth\ simulations are large enough to allow for such smoothing. Scales larger than 6$h^{-1}$ Mpc in radius would begin to incorporate information from the edges of the snapshots -- affected by projection and contamination effects -- into the sphere of interest surrounding the cluster. Therefore we limit this study to examining oriented structure at the smaller end of the typical cluster-cluster-bridge scale range, selecting a radius of 6$h^{-1}$ Mpc.


For this scale, we smooth the number density map with a Gaussian beam. A Gaussian is chosen rather than a top-hat filter due to the sparseness of galaxies; the top-hat filtered field exhibits sharp circular features while the continuous Gaussian-smoothed field allows for smooth computation of first and second derivatives. A Gaussian beam with a given full-width half-maximum (FWHM) incorporates similar information as a top-hat filter with radius $R_{\mathrm{TH}}$ when FWHM$\sim1.67 R_{\mathrm{TH}}$. Thus to include $\sim$6\hMpc\ worth of galaxy number density information on any side of the location of interest, we smooth with a 10\hMpc\ FWHM ($\sigma\sim4.25$\hMpc). To avoid unrealistic edge effects, we extend the galaxy number density $n_g$ arrays as zeros beyond the volume edges, assuming an empty background.

If $\tilde{n_g}$ is the smoothed map, the Hessian at any point in the map is defined as:
\begin{equation} \label{eq:Hessian}
H = \begin{bmatrix}
\frac{\partial^2 \tilde{n_g}}{\partial x^2} & \frac{\partial^2 \tilde{n_g}}{\partial x\partial y} \\
\frac{\partial^2 \tilde{n_g}}{\partial y\partial x}  &\frac{\partial^2 \tilde{n_g}}{\partial y^2}
\end{bmatrix}
\end{equation}
Following the conventions of \citet{BE1987} who studied the properties of peaks in the CMB,
we flip the sign of the Hessian eigenvalues $\lambda_i$ such that they are defined to be positive at peaks and negative at troughs. The eigenvalues are ordered as $|{\lambda_1}|>|{\lambda_2}|$, such that $\lambda_2$ and its corresponding eigenvector $\boldsymbol{v_2}$ describe the axis of slowest change in curvature, i.e., the long-axis of structure. $\boldsymbol{v_2}$ has a rotation angle $\theta$ from the $x$-axis. Each $y$ map is rotated by $-\theta$ to align $\boldsymbol{v_2}$ with the $x$-axis. Figure~\ref{fig:orientation} shows examples of several orientations determined for the smoothed galaxy map. After rotation, each map is stacked such that the long-axis / filament axis is aligned throughout.

\section{Results}\label{sec:results}

Oriented stacks of $y$ for splits of halo and diffuse gas using $R_{200c}$, averaging over 98 snapshots, are shown in Fig.~\ref{fig:allstacks_r200} in logarithmic scale. The $y$ contribution from the warm-hot and hot gas is also shown. The cool-gas $y$ map is not shown as it the stacked contribution from particles with $T<10^5$~K is $y\sim2\times10^{-9}$ at maximum, negligible in comparison to all other categories. The halo gas map shows a strong central signal from the stacked central cluster with $M>10^{14}$\hMsun; this is also seen in the y map of only massive clusters. The diffuse gas map, made up of an entirely distinct set of particles as the halo gas map, also has a non-zero central signal surrounded by a ring of higher $y$. This morphology comes from averaging over shells of high-pressure gas just beyond $R_{200c}$ of the central stacked clusters. Furthermore, careful examination reveals many such shells of $y$ further afield in the diffuse map, whose locations correspond to halos in the massive and low-mass cluster maps. The association between halos and diffuse gas is further explored in Sec.~\ref{subsec:halo_assoc}.

The stronger signal along the horizontal filament axis is \RH{visible} in each stack, however it appears more concentrated for the halo gas and hot gas maps than for the diffuse and warm-hot maps. This indicates that the diffuse gas is less compressed along the filament axis defined by the galaxies. The puffier appearance relates to the fact that energy injected by AGN sphericalises at large distances from the galaxies \citep{Dave2019, Cui2022}.

To quantify these observations, we also decompose each stack $I$ into multipole moments $m$:
\begin{equation}
    I(\theta,r) = \sum_m \left (  C_m(r) \cos(m \theta)+S_m(r) \sin(m \theta) \right ),
\end{equation}
where $r$ and $\theta$ are the polar coordinates. We will focus only on the first two even cosine amplitudes, $C_0$ and $C_2$. The $C_0$ monopole term quantifies the isotropic profile of the gas, equivalent to the circularly-averaged profile of an unoriented stack. The quadrupole $C_2$ quantifies most of the anisotropy of the gas, summing the signal from an opening angle around the $x$ axis. $C_2$ is useful especially when comparing observational oriented stacks to expectations from simulations, as it depends on not only the gas processes in the simulation but also the shapes and extents of large-scale structures like filaments, which hold cosmological information. Higher $C_m$ moments are cosmologically interesting (see L22), but \RH{have lower SNR. The key questions of this study can be addressed without these higher order moments.} 
$S_m(r)$ quantifies the contribution along the vertical misaligned axis. It may provide information about underdense regions, \RH{but we leave this to future work.}

The radial profile of the cosine components are taken by
\begin{equation}\label{eq:multipole_moments}
    C_m(r) = \frac{1}{X\pi}\int_0^{2\pi} d\theta I(\theta,r) \cos{(m \theta)},
\end{equation}
where $X=2$ when $m=0$ and $X=1$ when $m=2$. To quantify the variance in $C_m(r)$ among the 98 snapshots, we perform the decomposition for each individual rotated $y$ map from each snapshot. The standard error on the mean (SEM) for each bin in each multipole profile is the standard deviation of the bin across the 98 independent maps divided by $\sqrt{98}$. In the resulting figures, error bars show $\pm$1 SEM. We examine results only out to 8\hMpc\ from the central cluster to avoid contamination and projection effects in the snapshot edges.

In the following sections, the discussion focuses on addressing the key questions for modelling the anisotropic tSZ signal in filaments and superclusters. The first addresses \textit{extent}--to what radius should a halo model go to capture a sufficient fraction of the $y$ signal?--and the second addresses \textit{mass range}--is it more important to understand the pressure profiles for halos in a particular mass range than others? 

\subsection{Radial contributions -- Monopole}

\begin{figure}
    \centering
    \includegraphics[width=\columnwidth]{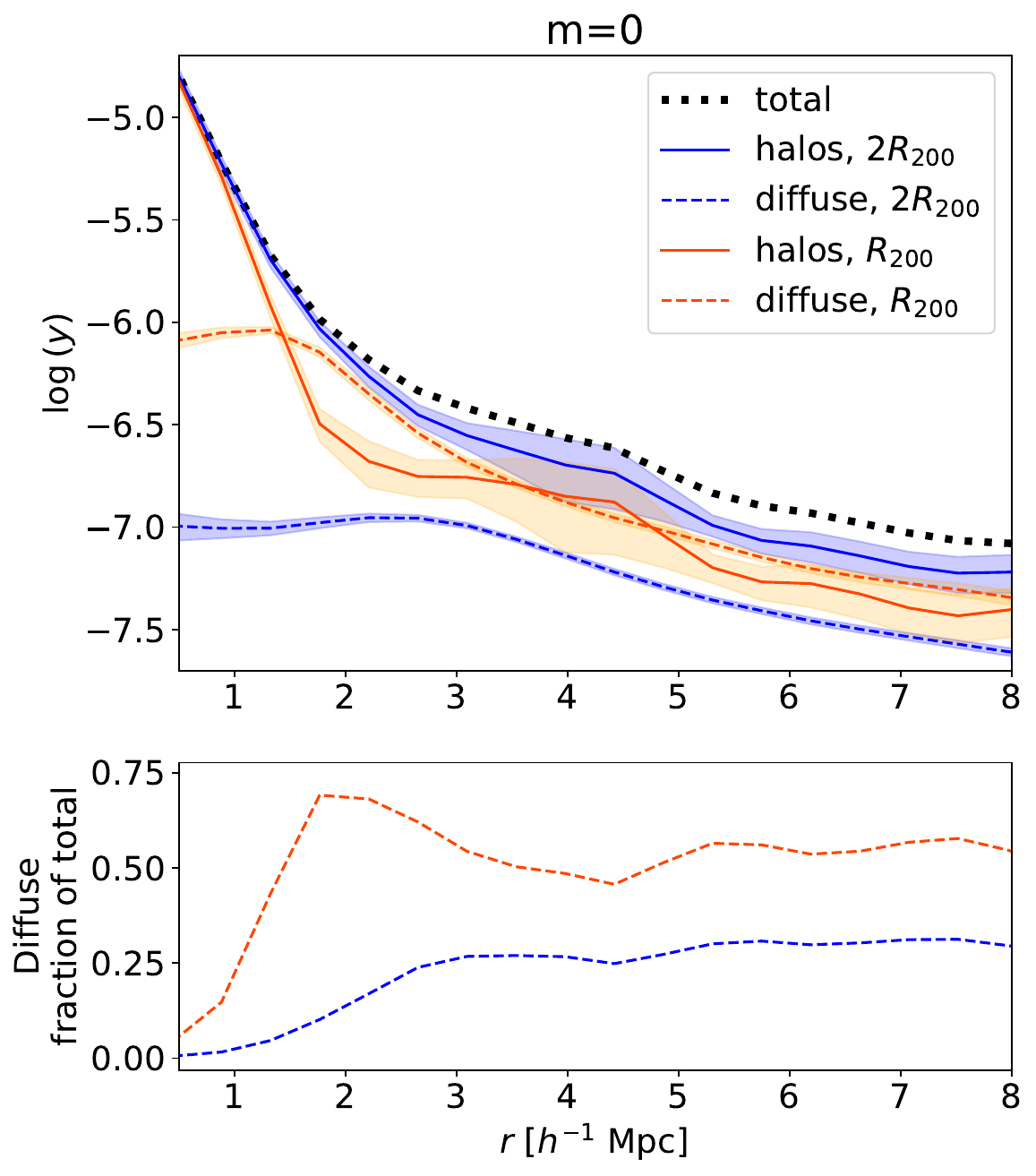}
    \caption{Monopole moment of the stacked $y$ maps for halo gas (solid) vs. diffuse gas (dashed) particles divided by $R_{200c}$ (orange) and by $2R_{200c}$ (blue). The signal from all particles is shown in black. Shaded error regions encompass $\pm$ 1 SEM. Errors are not shown for the total curve for visual clarity. The lower panel shows the fraction of total $y$ that the diffuse gas contributes in each case.}
    \label{fig:monopole}
\end{figure}

The angle-averaged $y$ profile is shown in Fig.~\ref{fig:monopole} for halos versus diffuse gas using $R_{200c}$ (orange) and $2R_{200c}$ (blue) as the boundary. By construction, the halo gas dominates near the centre in both cases ($y\gtrsim10^{-5}$; the exact values in the cluster centre depend on the binning). The diffuse gas from beyond $R_{200c}$ is over an order of magnitude lower ($y\sim10^{-6}$), but non-zero, at the centre because of projection of $y$ signal from a shell of gas beyond the threshold radius. \RH{As the dividing line between halo/diffuse gas is pushed to $2R_{200c}$, the amplitude at the centre of the diffuse} component decreases by another order of magnitude. For $R_{200c}$, as the profiles reach $\sim1.5$~\hMpc, the contribution from diffuse gas begins to dominate the central stacked cluster outskirts. Further afield, the $y$ signals from halo and diffuse gas are comparable: the gas beyond $R_{200c}$ contributes $\sim$half or more of the stacked tSZ signal far from clusters.

\begin{figure*}
    \centering
    \includegraphics[width=\textwidth, trim={0cm 0cm 0cm 1cm},clip]{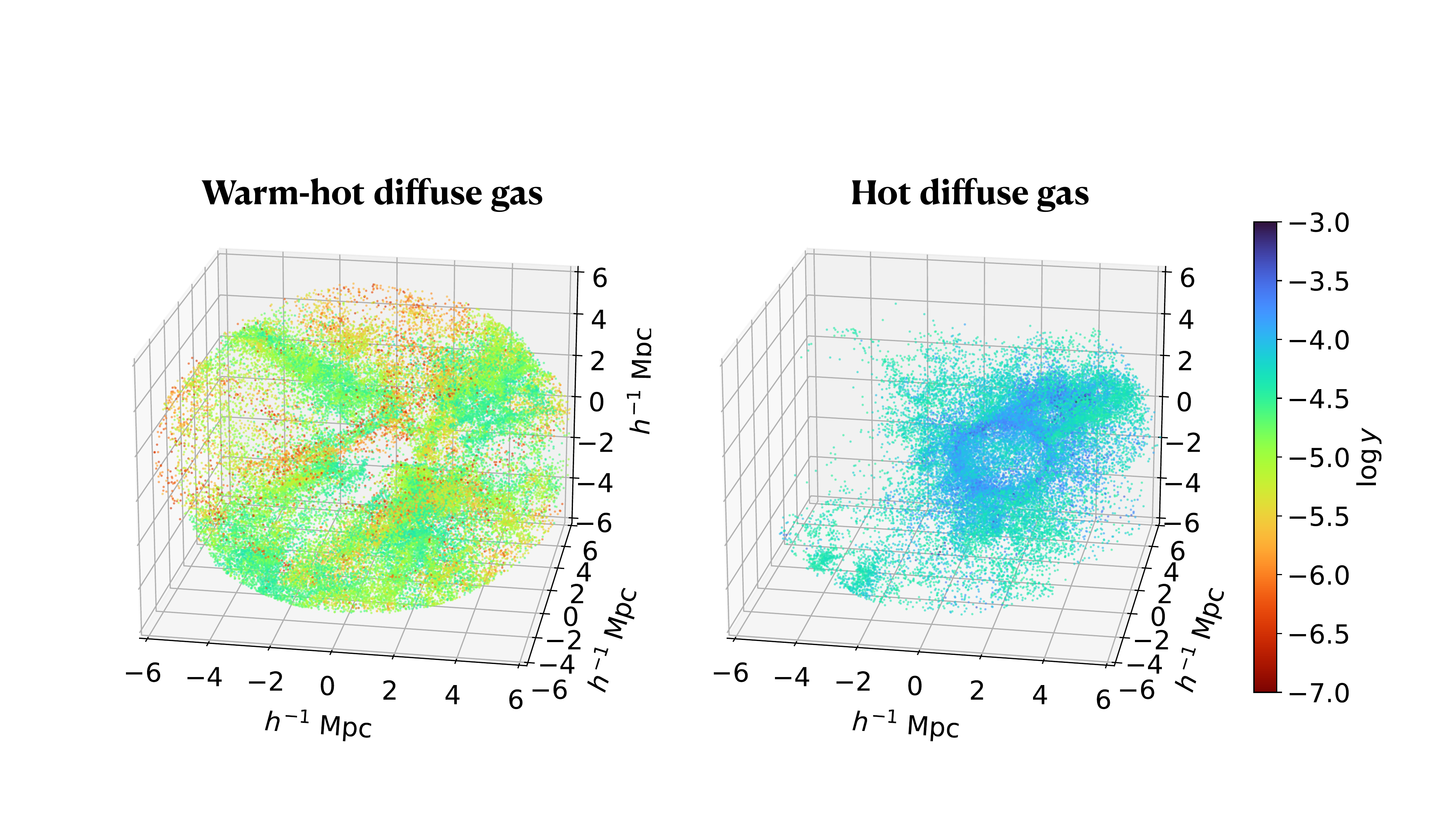}
    \caption{Representation of the 3-dimensional structure of the warm-hot ($10^5<T<10^7$~K, left) and hot ($T>10^7$~K, right) gas beyond $R_{200c}$ of the halo centres. The snapshot is the same as that shown in Fig.~\ref{fig:orientation} (a), with similar orientation. The hot gas dominates the $y$ signal along the filament axis. Note that the $y$ values for individual particles shown here are higher than typical values in the projected $y$ maps due to the smoothing that is applied when making the maps.}
    \label{fig:gas_division_yi}
\end{figure*}

The comparison is more distinct when $2R_{200c}$ denotes the threshold between halo and diffuse gas. As described above, the central $y$ signal from projected diffuse gas is several orders of magnitude below the halo signal, and an order of magnitude lower than the $R_{200c}$ case. This order-of-magnitude decline, from excluding gas particles in a $1R_{200c} < R <  2R_{200c}$ shell surrounding the central stacked clusters from the diffuse category, speaks to the non-negligible pressure of particles in that shell. We further investigate the pressure and distribution of these particles in Sec.~\ref{subsec:halo_assoc}. Further out in the oriented stack, although the gas signal from neighboring halos continues to dominate the signal, the diffuse gas fraction rises and contributes a significant fraction of the total: $\sim25$\% beyond $r\sim3$ \hMpc.

The diffuse gas beyond $R_{200c}$ is consistently higher than the $2R_{200c}$ curve, due to the $y$ contributions from gas which lies between $R_{200c} < R < 2R_{200c}$ of halos in the far-field from the central cluster.

\subsection{Association with halos}\label{subsec:halo_assoc}

Multiple physical processes are responsible for heating the diffuse gas before $z=1$. Gas is shock-heated as it falls into filaments during the earlier formation of the cosmic web; later, it is shock-heated from infall into halos. Conversely, gas which has fallen into halos is ejected by processes like AGN feedback. We briefly investigate how associated the diffuse gas is with halos, and what its likely heating source is, at $z=1$.

To address this, we examine the 3D distribution of the individual $y_i$ contributions of each $i$ gas particle in a single simulation snapshot at $z=1$. Fig.~\ref{fig:gas_division_yi} shows the same snapshot as Fig.~\ref{fig:orientation} (a), with a similar orientation; the filament axis identified using galaxies runs from bottom-left to top-right. When selecting for diffuse gas beyond only $R_{200c}$, gas with temperatures above $10^7$ K in the outskirts of massive halos dominates the tSZ signal. This gas, strongly correlated with halo positions, is thus highly correlated with the galaxies and preferentially lies along the filament axis.

When the threshold is extended to $2R_{200c}$, the remaining gas is more diffuse, less hot, and visually appears less associated with halos (not shown). However, a brief investigation shows that higher-pressure shells surrounding some halos are still visible beyond $2R_{200c}$ and even out to $4R_{200c}$. This visual evidence is supported by examining the average gas pressure in shells of increasing radius ($1-2R_{200c}$, $2-3R_{200c}$, etc) from the most massive halo centre in various snapshots. The average pressure continues to decline beyond 2$R_{200c}$. However, given that the central massive cluster radius is $\sim1-2$~\hMpc, excluding particles to larger radii than $\sim3R_{200c}$ begins to encroach upon the simulation's untrustworthy edge region. A more thorough study of the pressure profile to larger radii is beyond the scope of this work and likely better suited to a larger-volume simulation, but this perfunctory analysis suggests that when attempting to model the tSZ with a combination of halo component and field component, the boundary should be $\sim 4R_{200c}$ or even larger.

Next, we visually examine the direction of gas velocities for the shells of gas around $R_{200c}$ of the most massive halos in the simulation. We find that in some snapshots, the gas is primarily infalling, while in others the trajectories are mostly outgoing. This indicates that there is both shock-heated infall and AGN-heated outbursts of gas contributing to the associations of the diffuse gas with halos. A more detailed investigation of the history of the gas is beyond the scope of this work, but interested readers may further examine the dynamic nature of the gas by viewing videos provided by \tth\ collaboration\footnote{\url{https://music.ft.uam.es/videos/music-planck}}.

These findings are similar to a recent study of multi-phase gas $z=0$ in the Illustris TNG simulations \citep{Gouin2022}, which found that hot gas extends well beyond $R_{200c}$ for galaxy clusters and that both the hot and warm-hot phases beyond $R_{200c}$ include both infalling and outgoing gas.

\subsection{Anisotropy -- Quadrupole}

The monopole results indicate that modelling the diffuse gas well beyond the halo virial radius is important. However, observationally, the quadrupole and other higher-order moments offer several advantages. Uncorrelated projection effects and long-wavelength contamination from the CMB are two systematics that enter into cluster tSZ stacks. Both are isotropic in a stack of sufficient numbers of clusters. Thus, their contribution must be estimated and subtracted to make a measurement for the monopole -- especially far afield where the contamination becomes comparable or greater than the signal. \RH{Critically,} the quadrupole of an oriented stack is insensitive to these isotropic systematics. Additionally, it has SNR benefits from the stacking of extended structure. The remainder of the paper focuses on the quadrupole moment \RH{, rather than the monopole}, to provide guidance for future observational work.

The quadrupole result for the same categories as in Fig.~\ref{fig:monopole} tells a similar story as the monopole as to how extending the radius from $1-2R_{200c}$ affects the $y$ signal; thus we elect not to show it. The quadrupole, however, adds information about the anisotropy of the diffuse gas: near the stack centre, the anisotropy is small and the $m=2$ signal is an order of magnitude below the $m=0$ signal. However, further afield ($\gtrsim 4$~\hMpc), the distribution becomes more anisotropic, with $m=2$ becoming similar to $m=0$ in magnitude. The increase in anisotropy moving out from the stack center can be seen by eye in the diffuse gas map of Fig.~\ref{fig:allstacks_r200}.

\begin{figure}
    \centering
    \includegraphics[width=\columnwidth]{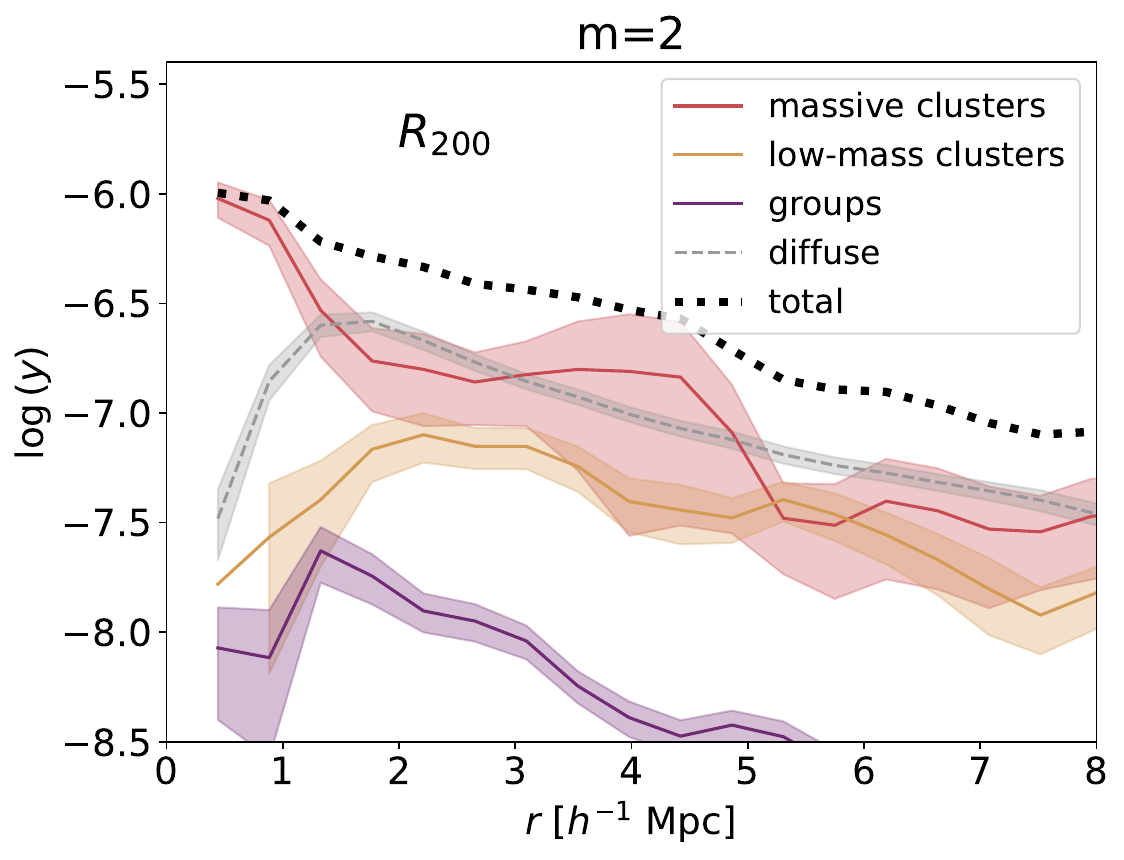}
     \includegraphics[width=\columnwidth]{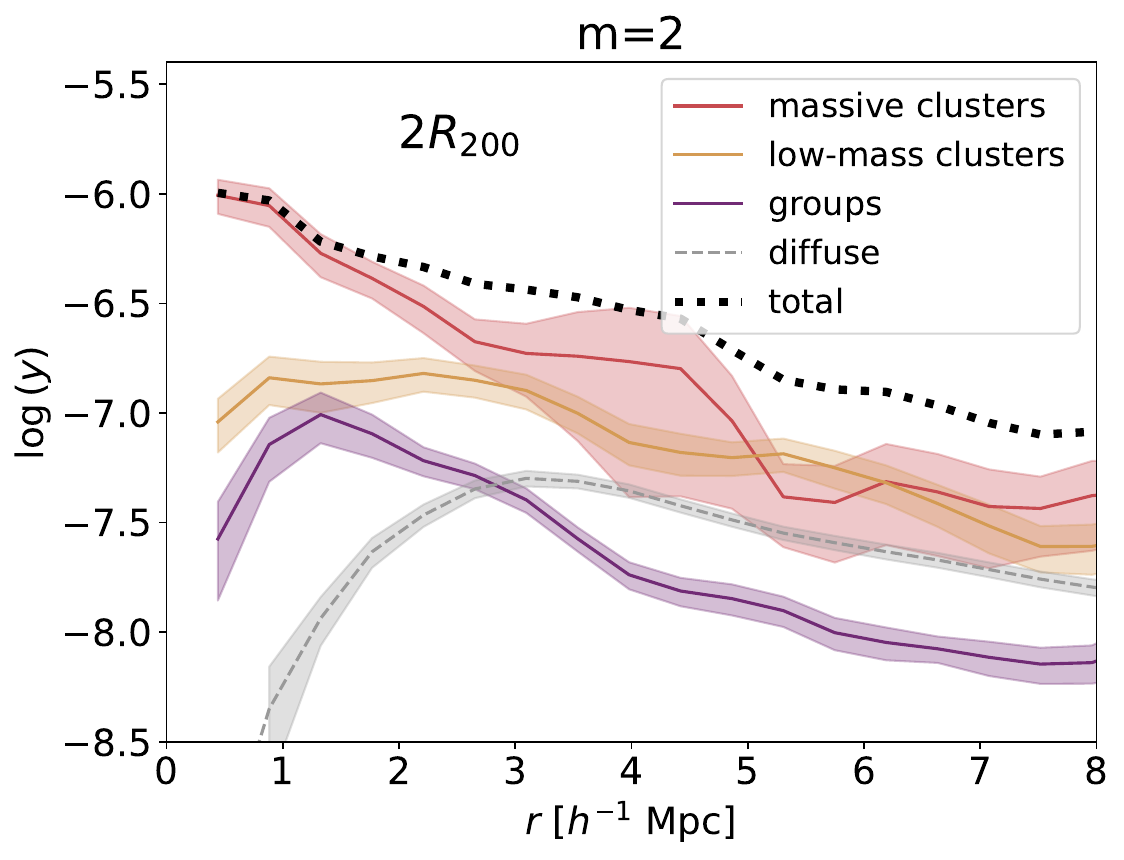}
    \caption{Quadrupole of the stacked $y$ maps for gas associated with different halo mass ranges by $R_{200c}$ (above) and $2R_{200c}$ (below). Note that there is significant overlap between particles in different halo categories in the $2R_{200c}$ case. The boosts in $y$ from all halo categories when going from 1 to 2$R_{200c}$ reveals the importance of gas in the outskirts of halos.}
    \label{fig:mass_ranges_y}
\end{figure}

Having considered the question of extent, we turn to mass range. Fig. \ref{fig:mass_ranges_y} (top) shows the $m=2$ moments of the stacked images for the case where $R_{\mathrm{200c}}$ defines the halo/diffuse threshold. Among the halo gas categories, unsurprisingly, the high-mass clusters dominate. Beyond $2~h^{-1}$ Mpc, low-mass clusters cause a $y$ signal in the quadrupole roughly $\sim2/3$ as strong as the signal from high-mass clusters. Groups contribute very little to the quadrupole signal, indicating that it is most important to accurately model the signals from halos above $M=10^{13}$\hMsun, in order to extract the $y$ signal from gas beyond $R_{200c}$ in observational data. 

The narrative, however, is markedly different if the threshold is extended to $2 R_{200c}$ (see the bottom panel of Fig.~\ref{fig:mass_ranges_y}). Adjusting this threshold adds gas pressure to the halo categories and removes it from the diffuse category. All halos experience some boost in $y$, with groups exhibiting the largest difference; meanwhile, the diffuse gas signal is depleted. We caution that some of the group boost effect comes from overlap with neighboring halos -- as the halo radius extends out to $2R_{200c}$ from group centres, some of the spheres begin to intersect with the outskirts of higher-mass halos. In such cases, it is difficult to assess whether a particle should truly belong to one category or the other, so we choose to allow the same particle to be assigned to multiple categories. $\sim11\%$ of group particles are also in the low-mass cluster category for $2R_{200c}$. Although this explains some of the disproportionate group $y$ increase, it may also be due to the halo-mass-dependent effects of AGN feedback. The general understanding is that AGN feedback is most effective at altering the pressure profiles of low-mass halos, depleting gas from the centres and moving it beyond the virial radius \citep{Sorini2022}, causing shallower pressure profiles as compared to predictions from self-similarity \citep[e.g.,][]{Hill2018}.

Whether extending to 1 or $2R_{200c}$, the diffuse gas is important beyond $1-3$~\hMpc. From the perspective of cosmological modeling of the extended tSZ, then, the key takeaway from these findings is that even accurate modelling of pressure profiles out to 2$R_{200c}$ and down to $M=10^{12}$\hMsun\ is insufficient for making precise cosmological inferences from oriented stacks. A well-tested model for the diffuse gas is needed.

\begin{figure}
    \centering
    \includegraphics[width=\columnwidth]{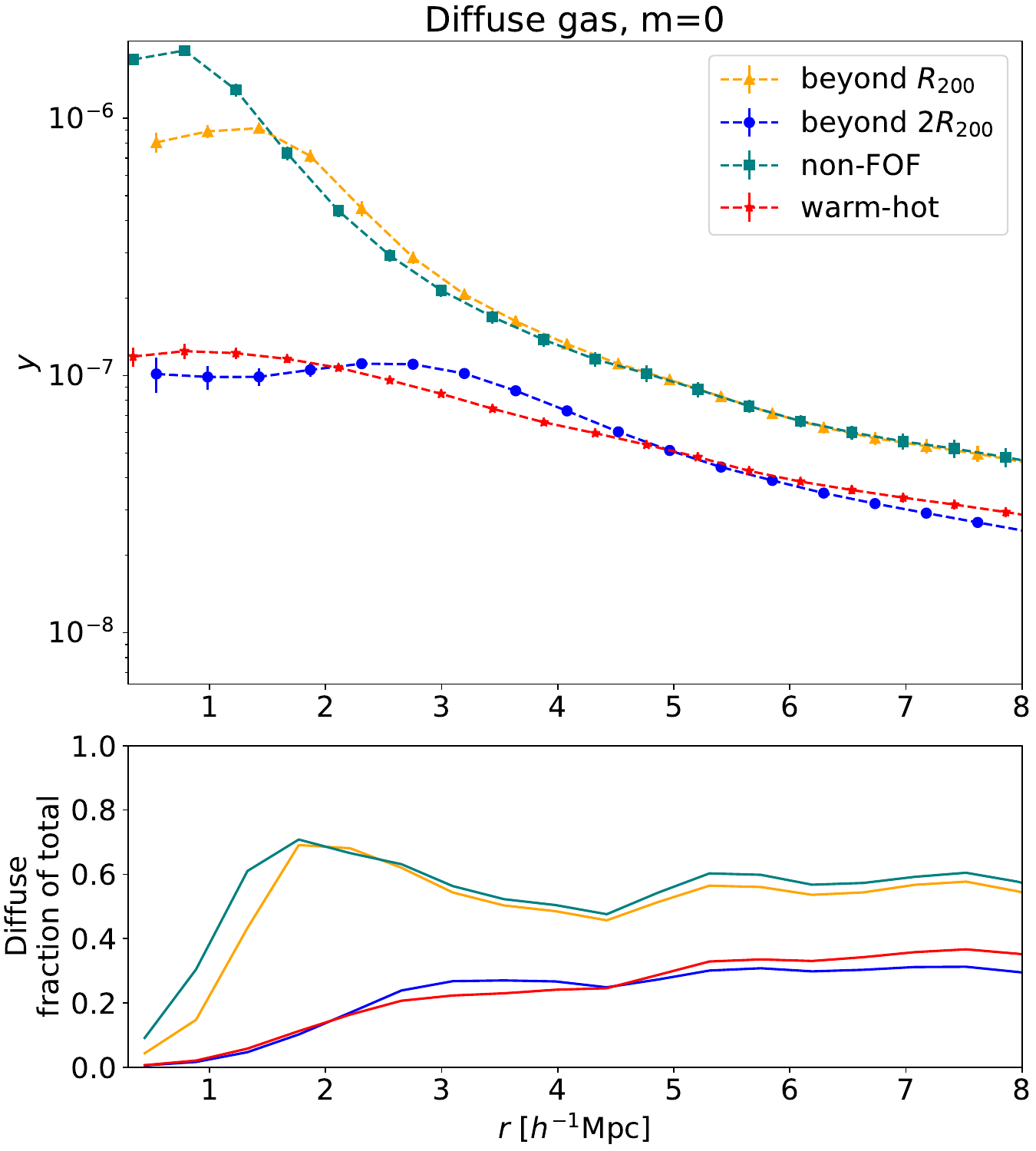}
    \includegraphics[width=\columnwidth]{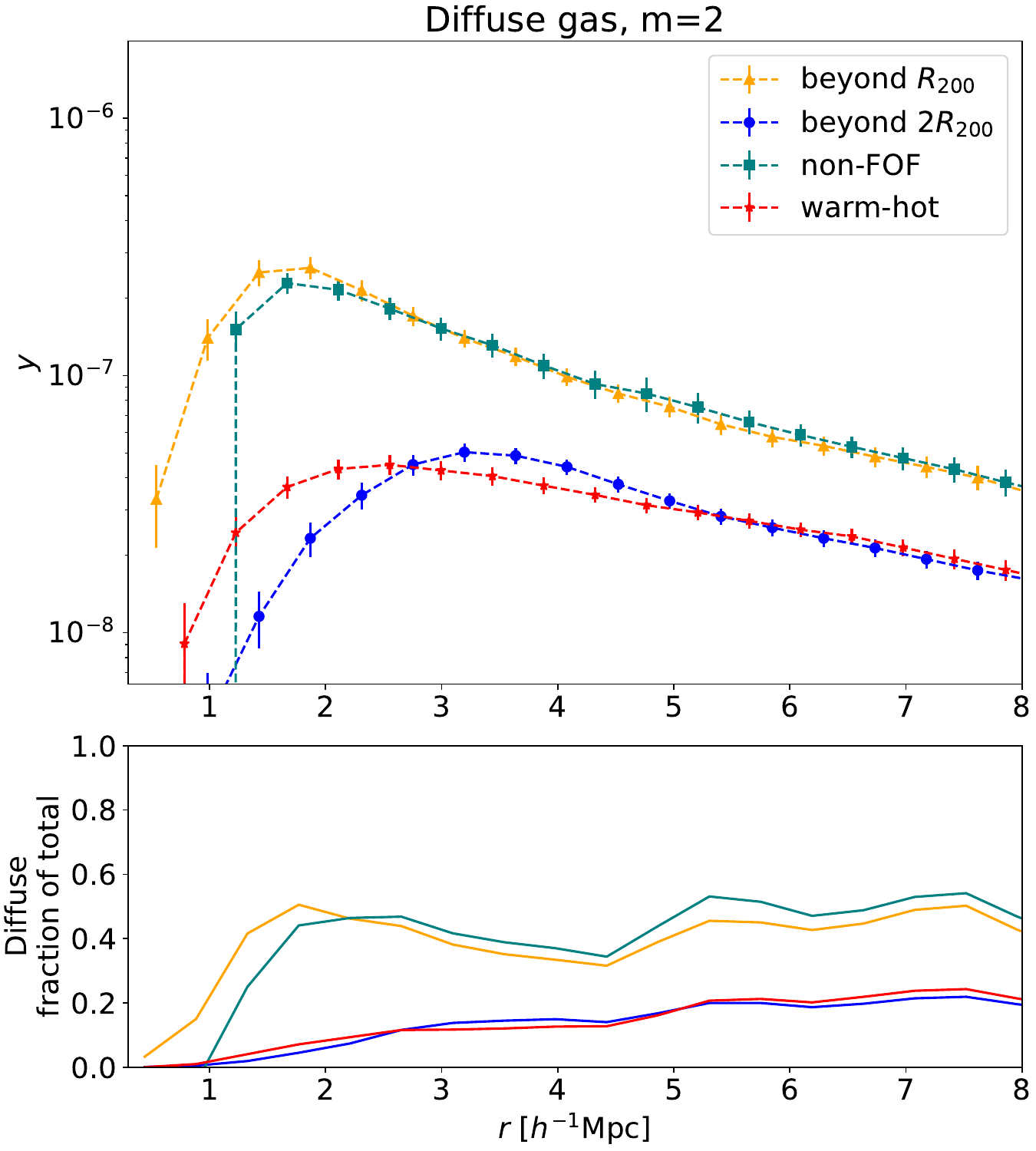}
    \caption{Monopole and quadrupole decompositions of the diffuse-gas stacks for the four different definitions. The lower panel shows the fraction of the total $y$ signal in that multipole that the diffuse gas contributes. The curves are artificially offset in $r$ for visual distinction. In any definition, the contribution from diffuse gas far afield from the massive central halo is significant: $\sim20\%$ at minimum.}
    \label{fig:mult_unbound_frac}
\end{figure}

To better visualize the percentage that the diffuse gas contributes to the total oriented $y$ signal under different definitions, we plot the tSZ contribution from each definition in Fig~\ref{fig:mult_unbound_frac}. This figure repeats the orange and blue dashed lines from Fig.~\ref{fig:monopole} for $m=0$, while adding additional information in $m=0$ and also including $m=2$. In the isotropic profile, gas unassociated with FOF halos is the largest contributor near the central stacked cluster. This is likely due to the unusual halo shapes which FOF often defines; the massive central halo is more likely to be oriented along any other axis besides the line of sight, so along the line-of-sight there is more projection from non-FOF gas than in other cases. 

In both $m=0$ and $m=2$, the warm-hot gas and gas beyond $2R_{200c}$ give very similar results, generally contributing less than 50\% of the total $y$ signal. Their low amplitudes are a consequence of the fact that both categories are missing the hot gas which exists between 1 and 2 $R_{200c}$ from halo centres (as discussed in Sec.~\ref{subsec:halo_assoc}), which does contribute to the two other categories. The warm-hot and beyond-$2R_{200c}$ categories overlap to a large extent; nearly 80\% of the warm-hot gas lies beyond $2R_{200c}$ in the average snapshot. The remainder, which is mostly within groups, only slightly shifts the shape of the profiles with respect to the beyond-$2R_{200c}$ category.

Despite the differences in definition, in all methods, the diffuse gas is important. At minimum, it contributes $\sim20-30\%$ of the tSZ signal 6\hMpc\ from the central cluster, and at maximum, over 50\%.

\subsection{Axis comparison}
\begin{figure}
    \centering
    \includegraphics[width=\columnwidth]{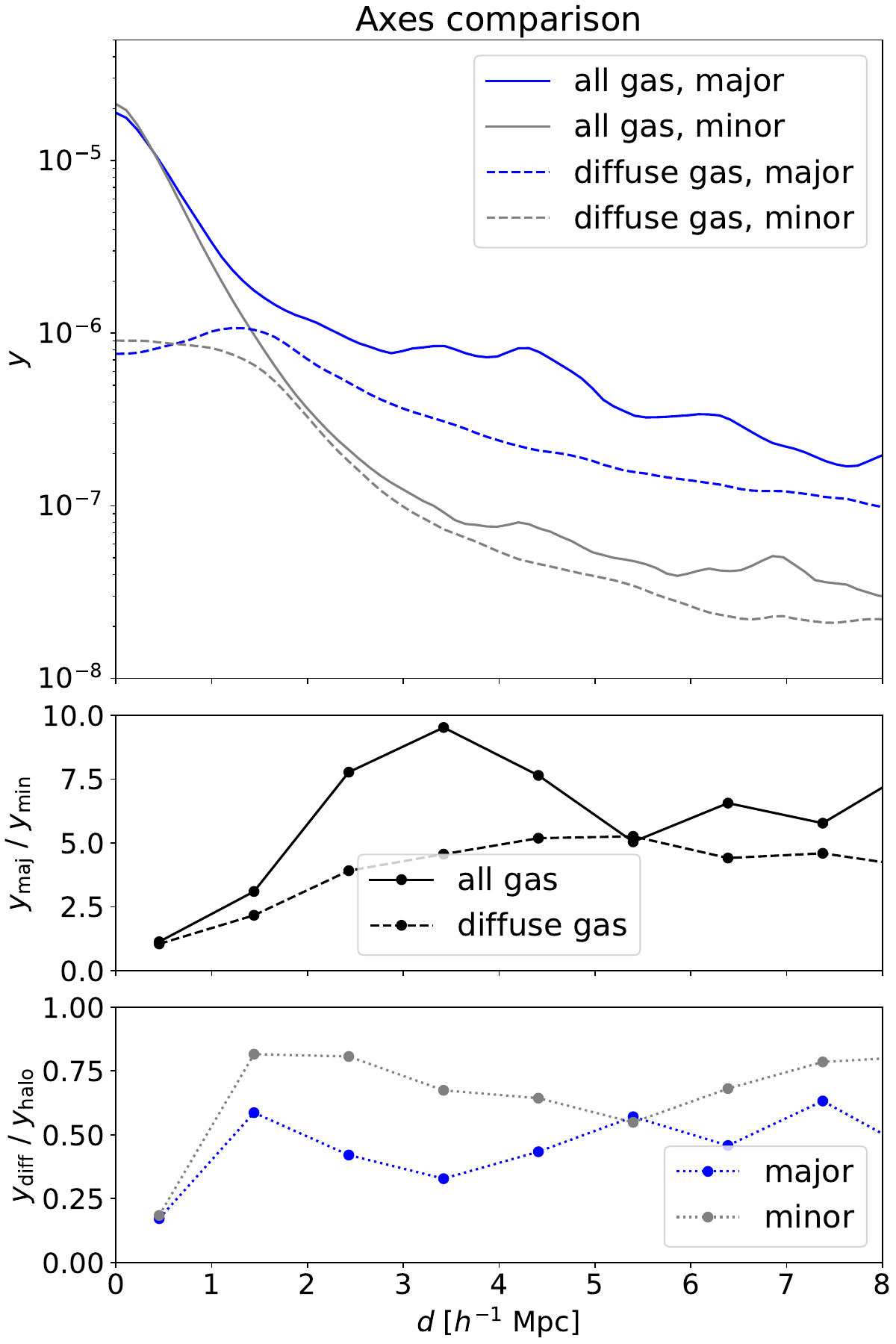}
    \caption{Top: comparisons between the $y$ signal along the major (blue) and minor (gray) axes of the stacked $y$ maps made from all gas particles (solid) and made from particles beyond $R_{200c}$ (dashed). Middle: the ratio of the major to minor axis for each gas category. Bottom: the ratio of diffuse $y$ to total $y$ for each axis.}
    \label{fig:xyaxis}
\end{figure}

Finally, to demonstrate the usefulness of oriented stacking, we quantify the differences along the major (filament) and minor (perpendicular to filament) axis profiles of the stacks. Fig.~\ref{fig:xyaxis} shows the profiles as a function of the major/minor axis coordinate $d$ when averaged over a rectangular band 5 pixels ($\sim1.5$ Mpc) wide, centred on each axis. This is distinct from doing a higher-order multipole decomposition. Shown in solid lines, the total $y$ profile along the major axis (blue) is higher than along the minor axis (gray). The middle panel shows the ratio of the signals, which ranges from 2-7 and generally increases with $d$. By construction, the major axis contains more gas associated with halos, which is much hotter and denser than average. For the diffuse gas alone (dashed), the gas along the major axis also contributes a stronger $y$ signal than the minor axis. This is likely due to a combination of the following reasons: the gas is denser and hotter due to collapse along the filament and towards halos; it has been heated by AGN feedback near halos; and there is more gas along the filament axis to contribute to each line-of-sight integral.

We note that, by orienting on filaments, the minor axis becomes depleted with respect to the average, which inflates the major-to-minor axis ratio. To determine how much oriented stacking boosts the signal compared to an unoriented stack, we compare the major-axis signal to the monopole signal, which represents the angle average (as shown in Fig.~\ref{fig:monopole}). Oriented stacking boosts the extended signal along the major axis by about $2\times$.

An interesting question is whether the diffuse gas contributes a higher or lower \textit{fraction} of the total $y$ signal along the filament versus off. The lower panel of Fig.~\ref{fig:xyaxis} addresses this; the fractional contribution is slightly higher, on average, along the minor axis. Although this may seem counterintuitive at first, it is unsurprising. The main difference between the axes is that the total mass of halos along any line of sight on the major axis is higher than along the minor axis. There is a known strong relationship between mass and halo gas pressure; the relation between integrated $Y$ and halo mass $M$ states that $Y\propto M^{5/3}$ (for massive halos). There is also expected to be some relationship between the diffuse gas pressure and higher overall halo masses, as larger halos have stronger AGN that can heat the surrounding gas, and gas is shock-heated as it falls onto the filament axis. However, Fig.~\ref{fig:xyaxis} suggests that the boost in diffuse $y$ signal along the filament is not strong enough to compensate for the boost in halo $y$ signal.

\section{Measures of superclustering} \label{sec:superclustering}

Lastly, we sort the 98 cluster regions by measures of their density and elongation, as determined by the galaxy field, to assess whether these measures are effective at increasing the signal from diffuse gas. The density and elongation of the large-scale galaxy environment are indicators of the overall superclustering of matter in a given region, and are both good predictors of the anisotropic $y$ signal in observations (L22). At the position of the massive halos selected in Sec.~\ref{sec:halo_sel}, we examine the Hessian of the smoothed galaxy number density $\tilde{n_g}$. Using the eigenvalues of the Hessian defined in Eq.~\ref{eq:Hessian}, the ellipticity $e$ is defined as:
 \begin{equation} \label{eq:e}
     e = \frac{\lambda_1 - \lambda_2}{2 (\lambda_1 + \lambda_2)}.
 \end{equation}
 
We measure this large-scale ellipticity $e$ and the large-scale density $\tilde{n_g}$ at the position of the most massive cluster and combine them to make a normalized measure of superclustering, $s$:
\begin{equation}
    s = \frac{e}{\langle e \rangle}+\frac{\tilde{n_g}}{\langle\tilde{n_g}\rangle,}
\end{equation} where the means ($\langle ... \rangle$) are taken over the sample of 98 snapshots. No further normalization or rescaling (e.g., dividing by the background field rms $\sigma$), is needed because the snapshots all take place at the same redshift.

It is expected that $s$ will enhance the filament signal from halo gas because of the strong correlation between galaxy number and gas pressure for massive halos. However, it is less obvious whether $s$ can also boost the signal from the most diffuse, unassociated gas. Figures \ref{fig:allstacks_r200} and \ref{fig:mult_unbound_frac} indicate that the warm-hot gas is very diffuse. As this gas is the object of missing baryon searches, we examine whether using $s$ to select regions of high superclustering is helpful for increasing the signal from the warm-hot gas.
\begin{figure}
    \centering
    \includegraphics[width=\columnwidth]{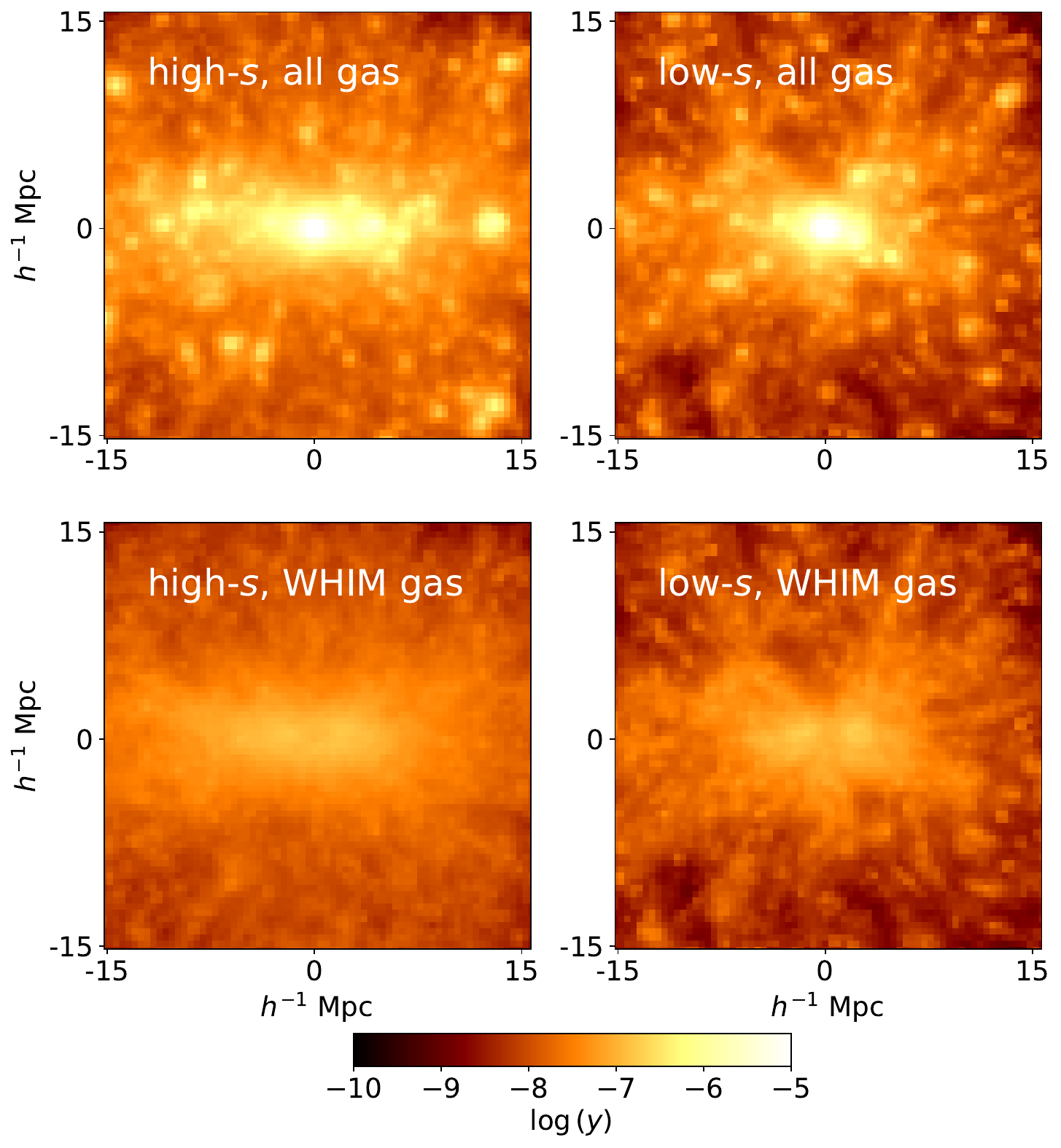}
    \caption{Stacks of the top 20 and bottom 20 snapshots, when rank-ordered by the measure of superclustering $s$.}
    \label{fig:slevels_ymaps}
\end{figure}

\begin{figure}
    \centering
    \includegraphics[width=\columnwidth]{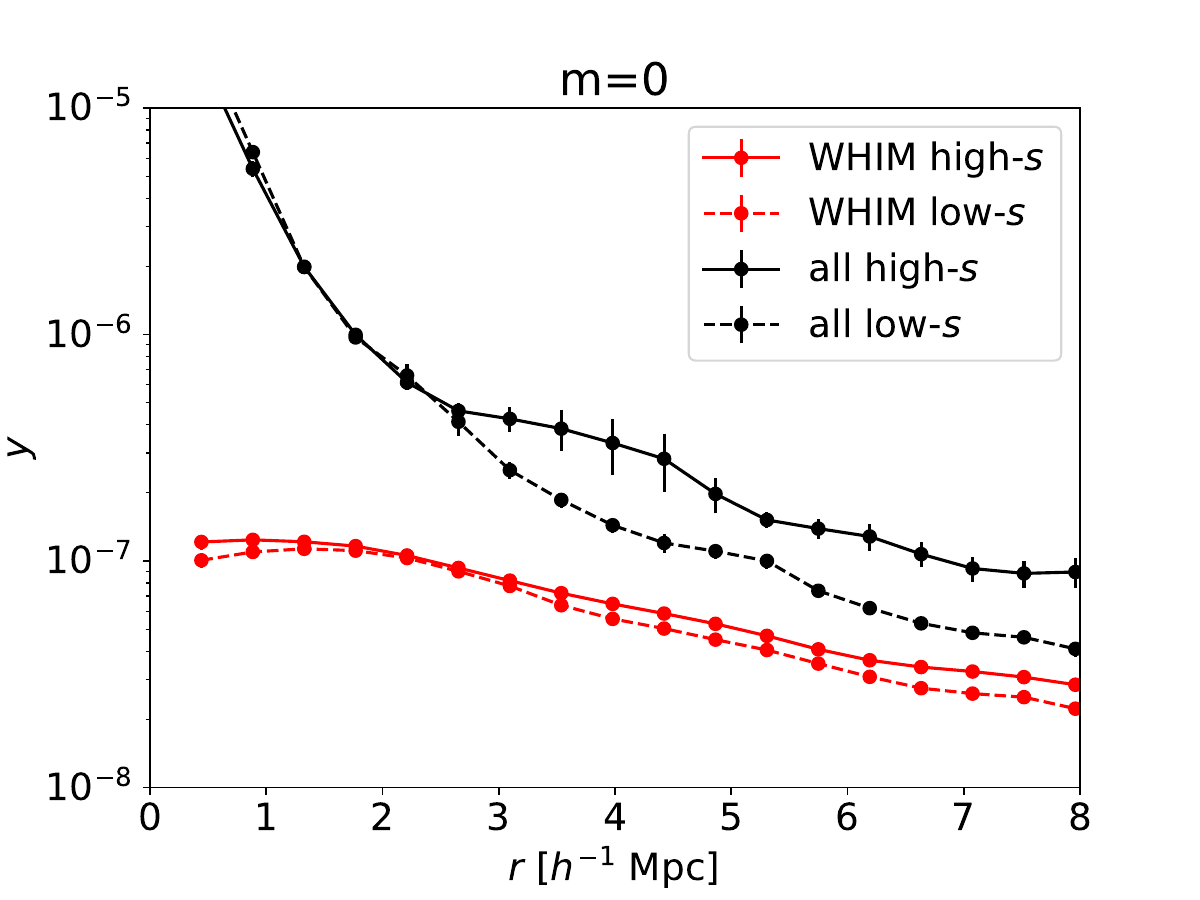}
    \includegraphics[width=\columnwidth]{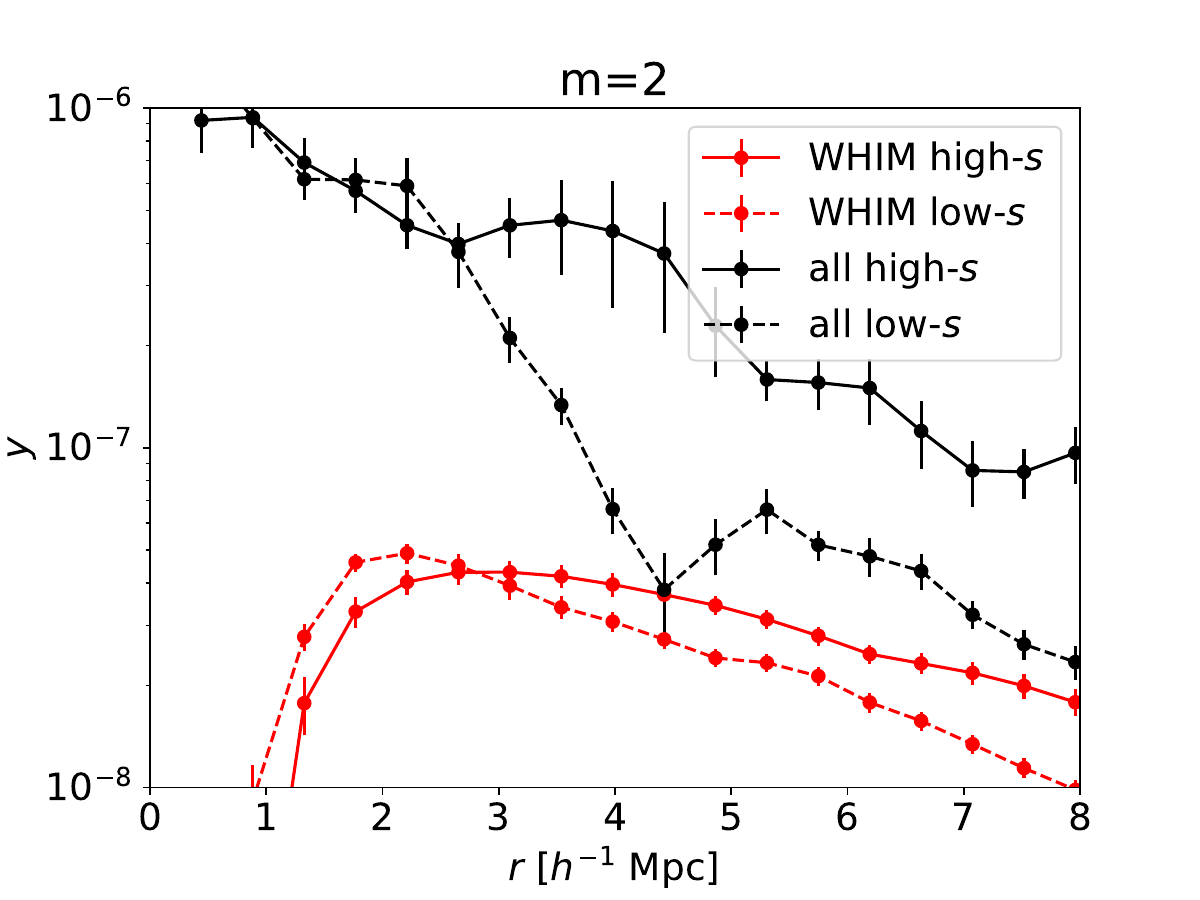}
    \caption{Monopole and quadrupole moments of the stacks shown in Fig.~\ref{fig:slevels_ymaps}. In both moments, far from the centre of the stacks, the high-$s$ regions exhibit stronger signal for warm-hot gas than low-$s$ regions.}
    \label{fig:slevels_multipoles}
\end{figure}

We rank-order the snapshots by $s$, then stack the rotated $y$ maps from the top 20 snapshots and bottom 20 snapshots. We also test samples of more or less than 20 to check for robustness. The results are shown in Fig.~\ref{fig:slevels_ymaps}. The effect of $s$ ranking has a clear impact in the all-gas signal, but the impact on the warm-hot gas is less clear. Fig.~\ref{fig:slevels_multipoles} quantifies the effects with a multipole decomposition. As expected, the $s$ measure has a higher effect on the total $y$ signal than it does on the warm-hot signal. Nevertheless, far beyond the central stacked cluster ($r\gtrsim 4$\hMpc, the high-$s$ warm-hot stacks show a subtle boost in $m=0$ (the warm-hot gas is $\sim10-30\%$ stronger) and a more distinct boost ($\sim30-70\%$) in $m=2$. Interestingly, in the near-outskirts of the central stacked cluster ($r\sim2$\hMpc), the effect in $m=2$ is opposite. It is unclear why the low-$s$ stack exhibits stronger anisotropy close to the central cluster.

In summary, this brief investigation demonstrates the potential of using large-scale environmental measures from galaxy catalogues to select for regions with higher diffuse gas content. In future, it would be interesting to investigate whether measuring the large-scale environment with different populations of galaxies (e.g., field galaxies versus cluster galaxies, divided by color) changes the correlation with diffuse gas. Such a study would aid in the observational search for the remainder of the missing baryons.

\section{Discussion and conclusions}\label{sec:conclusions}
In this work, we have applied oriented stacking to simulated maps of Compton-$y$ from \tth\ \simba\, runs. We have applied several different cuts to the gas particles to separate the particles associated with halos from the diffuse or unresolved particles. We focused our discussion on the separation that uses a sharp cutoff radius $R_{200c}$ or 2$R_{200c}$, because this is most readily applied observationally, either through masking or through halo prescription modelling.

We find that the contribution to the stacked filament tSZ signal from beyond $R_{200c}$ of all halos is approximately equal in magnitude to the contribution from gas within halos. However, much of this signal comes from the shells of gas between $R_{200c}$ and 2$R_{200c}$, indicating that the gas is still associated with halos despite being diffuse by this basic definition. The gas beyond this radius for massive and low-mass clusters is frequently hotter than $10^7$K, placing it outside of the warm-hot definition often used in simulations ($10^5\mathrm{K}<T<10^7\mathrm{K}$). This heating is likely due to both shocks on infall and AGN heating of outgoing gas, since gas velocities in the simulation show both ingoing and outgoing motions.

Thus, precise modelling of the gas from 1 to 2$R_{200c}$ is very important in order to use extended, stacked tSZ signals for cosmology. This conclusion is consistent with the recent observational study of galaxy stacking at slightly lower redshifts by \citet{Schaan2021}, which showed that a significant fraction of the baryons are beyond the virial radius. Meanwhile, \citet{Baxter2021} studied shocks in \tth\ simulations at $z\sim0.2$, finding a shock feature in the $y$ signal of relaxed clusters at several virial radii. Further studies of such shocks in simulations at varying redshifts, and observational studies of the outskirts of halos with coming data, will be important for more accurate modelling of the diffuse tSZ signal over a range of redshifts.

In our study, going out to $2R_{200c}$ captures 75\% of the signal; however, this is also insufficient for using anisotropic stacked tSZ signals for precision cosmology. Modelling the halo signal further out is a possible solution. For example, there appears to be some association of the gas pressure with halo locations even at $4R_{200c}$ and beyond. However, it is challenging to develop a prescription which extends so far due to the extensive overlap between neighboring profiles. In many overlapping cases, it is unclear which halo the gas should be assigned to. Furthermore, it is challenging to study such large halo extents in \tth\ simulations due to their limited volume. It would be useful for future work to examine the oriented tSZ signal beyond $4R_{200c}$ in larger-volume simulations, or attempt to separate the halo-associated gas with the field gas at all radii using non-spatial properties, to give insight into whether gas pressure modelling for the field is a necessary addition to halo modelling.

A caveat to this work is that in all definitions applied to separate halo from non-halo gas, halos smaller than $10^{12}$~\hMsun\ were treated as members of the diffuse category due to their low resolution. It is possible that in higher-resolution simulations, such halos (numerous as they are) would be associated with a significant fraction of the tSZ signal. An additional limitation is that our analysis was restricted to a single suite of simulations with one feedback prescription, and the results will likely vary depending on the feedback model. The small volume of \tth\ zoom simulations as compared to cosmological volumes make \tth\ well suited for re-runs with subtle changes to the feedback mechanisms, and future work will explore this.

We also demonstrated that environmental measures of superclustering, as determined by the galaxy field, augment the signals from the warm-hot diffuse gas. Such environmental measures can be applied to help detect this gas in existing data like that from ACT and \textit{Planck}, but claimed detections of unbound or diffuse baryons should clearly state the modeling that went in to account for halo gas.

Several previous works have claimed a detection of the WHIM with oriented tSZ data. To compare our simulated results with those observations, it is important to again emphasize the role that environment plays in the diffuse gas pressure. Even before imposing the additional constraints for superclustering, the full sample of simulations examined in this work are already constrained to rare environments. They host a rare massive cluster at $z=0$; at $z=1$, we are viewing a snapshot of dynamic, coalescing proto-cluster regions that have higher density and more activity than average. The diffuse $y$ signals in similar regions in the real Universe are likely to be higher than the average intergalactic filament.

To compare with observations, we examine the $y$ profiles from the warm-hot stacks along the filament axis (total signal; roughly equal to the sum of $y_{m=0}+y_{m=2}$ in the red curves in Fig.~\ref{fig:mult_unbound_frac}). Various groups have studied the low-$z$ cluster pre-merger bridge Abell 399-401, most recently \citet{Hincks2022} in ACT data, finding a $y$ signal from the bridge of $10^{-5}$. With the bridge being 4.4~\hMpc\ in projection, the centre is 2.2~\hMpc\ from each cluster; the warm-hot $y$ signal in our work is nearly two orders of magnitude smaller at the same distance. This is likely both because the A399-401 system is in a rare state where the inter-cluster gas has been heated prior to its impending merger, and possibly also related to the difference in redshift. On the other hand, both \citet{degraaff2019} and \citet{Tanimura2019} found a much smaller residual $y$ signal of $\sim10^{-8}$ between stacked pairs of massive SDSS galaxies from $z\sim0.4-0.75$, after subtracting the estimated contributions from the stacked halos, and the former group estimated that $y\sim(5-6)\times10^{-9}$ comes from diffuse gas. Based on the pair separation in those studies, the signal examined was at a distance of $3-7$~\hMpc\ from the halos; \tth\ warm-hot signal is $\sim10\times$ larger at that same distance. This may also be due to environmental differences--while the pair condition selects for overdense regions, it does not guarantee them to be in such dense proto-cluster environments as examined in this study. However, it may also be a hint that the feedback in \tth\ is too powerful in heating diffuse gas.

The setup in this work is most similar to that in L22, which did oriented stacking on DES optically-selected red-sequence clusters with an average halo mass of roughly $5\times10^{13}$~\Msun. That work did not attempt to separate the halo gas from non-halo gas. The orientations in that work were determined on similar scales as the current work ($\sim12$~\hMpc\ compared to $10$~\hMpc), and the corresponding aligned $y_{m=2}$ signal was found to be $\sim2.7\times10^{-8}$ at 7~\hMpc\ from the cluster. The $m=2$ component of the all-gas $y$ signal found for \theth\ is similar: only $\sim3\times$ higher. Again, this difference may be due to environment. In L22 it was found that when the DES sample is limited to only those clusters in large scale high-superclustering regions, the signal grows to values very similar to those found in the present work. Some differences may also come from observational limitations: galaxy photometric redshifts, a larger redshift bin size, and galaxy magnitude limits all contribute to orientation inaccuracy in real data which can bias the signal lower compared to the idealized simulation result.

Given the challenges in exactly replicating the environmental selections in existing filament-gas observational studies with \tth, it is not possible to comment decisively on whether these simulations produce diffuse gas in accurate amounts and thermal states. The range of values analyzed in this work fall within the ranges seen in observations thus far, and a qualitative description in which environment plays an important role in properties of the WHIM does well at encompassing the results. Future simulations encompassing a larger range of environments would be helpful to complement upcoming advances in observational data, so that anisotropic studies of the tSZ can be used to make quantitative constraints on feedback models.

Beyond the massive clusters, the level of $y$ signals in this work are not detectable in an individual tSZ image in any current data, or even future data, due to the noise levels. Appendix~\ref{sec:appendix} provides some examples of what $y$ maps from the simulations would look like given forecasted noise for the upcoming Simons Observatory \citep[SO,][]{SO2019} and CMB-Stage 4 \citep[CMB-S4,][]{CMBS42022}. Even with the improvements those surveys will bring, stacking will be necessary to increase the SNR sufficiently to measure the diffuse gas in all but the rarest systems. With current data, the diffuse gas is on the cusp of detectability with the methods employed in this work. At a level of $y\sim 10^{-7}$, the signal is of similar magnitude as the $1\sigma$ uncertainties in the oriented stacking analysis of ACT data (L22), which stacked $y$ map cutouts surrounding 5,500 DES clusters. To achieve a 3$\sigma$ measurement of a $y= 10^{-7}$ signal with ACT data, 9$\times$ more regions---with signals similar to \tth\ snapshots assessed in this work---are needed. Despite recent large increases in the overlap between the ACT and DES footprints, enabling future use of the full DES Y3 cluster catalog ($\sim10\times$ more clusters), most of these clusters are at lower masses than \tth\ clusters and thus their filamentary environments tend to produce smaller average tSZ signals. Therefore it is unclear whether the diffuse signal will be readily detectable with this method in incoming ACT data. However, with improvements to the methodology and/or use of upcoming SO data, the detection should be possible \textit{given accurate subtraction of halo gas signals,} a major challenge to any missing baryons search.

With the coming advances in galaxy and CMB data, improvements in modelling gas both within and beyond halos are needed to fully realize the potential of the tSZ as a cosmological probe. The analysis presented in this work provides guidance to future observation and modelling efforts regarding the elusive boundless baryons. 

\section*{Acknowledgements}
The authors thank the anonymous referee for a thorough and helpful review of the manuscript.

This work has been made possible by the `The Three Hundred’ collaboration\footnote{\url{https://www.the300-project.org}}, which has received financial support from the European Union's Horizon 2020 Research and Innovation programme under the Marie Sklodowskaw-Curie grant agreement number 734374, i.e. the LACEGAL project. The simulations used in this paper have been performed in the MareNostrum Supercomputer at the Barcelona Supercomputing centre, thanks to CPU time granted by the Red Espa$\tilde{\rm n}$ola de Supercomputaci$\acute{\rm o}$n. The CosmoSim database used in this paper is a service by the Leibniz-Institute for Astrophysics Potsdam (AIP). The MultiDark database was developed in cooperation with the Spanish MultiDark Consolider Project CSD2009-00064.

ML acknowledges support from the Natural Sciences and Engineering Research Council of Canada Postgraduate Scholarships - Doctoral. ML also thanks Karen Scora for coming up with the alliterative title.

WC is supported by the STFC AGP Grant ST/V000594/1 and the Atracci\'{o}n de Talento Contract no. 2020-T1/TIC-19882 granted by the Comunidad de Madrid in Spain. He also thanks the Ministerio de Ciencia e Innovación (Spain) for financial support under Project grant PID2021-122603NB-C21. He further acknowledges the science research grants from the China Manned Space Project with NO. CMS-CSST-2021-A01 and CMS-CSST-2021-B01.

JRB's research was funded by the Natural Sciences and Engineering Research Council of Canada Discovery Grant Program and a fellowship from the Canadian Institute for Advanced Research (CIFAR) Gravity and Extreme Universe program.

R.~H. is a CIFAR Azrieli Global Scholar, Gravity \& the Extreme Universe Program, 2019, and a 2020 Alfred P. Sloan Research Fellowship. RH is supported by Natural Sciences and Engineering Research Council of Canada Discovery Grant Program and the Connaught Fund. 
\section*{Data Availability}

The data underlying this work have been provided by The Three Hundred collaboration. The data may be shared on reasonable request to the corresponding author, with the permission of the collaboration.



\bibliographystyle{mnras}
\bibliography{oriented_whim_300} 




\appendix 
\section{Detectability} \label{sec:appendix}
\begin{figure*}
    \centering
    \includegraphics[width=.7\columnwidth, trim={0cm 0cm 0cm 2cm},clip]{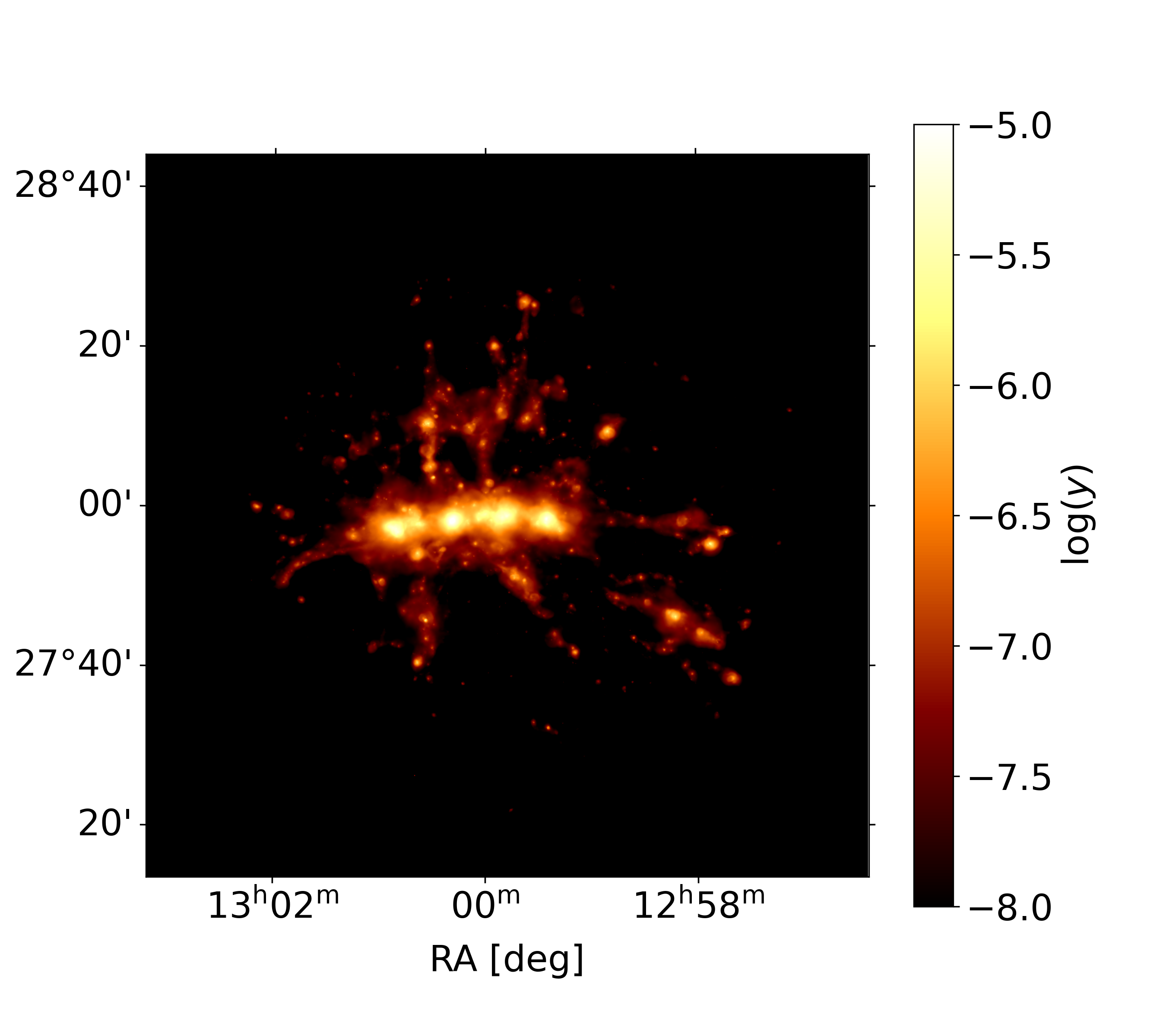} \includegraphics[width=1.3\columnwidth]{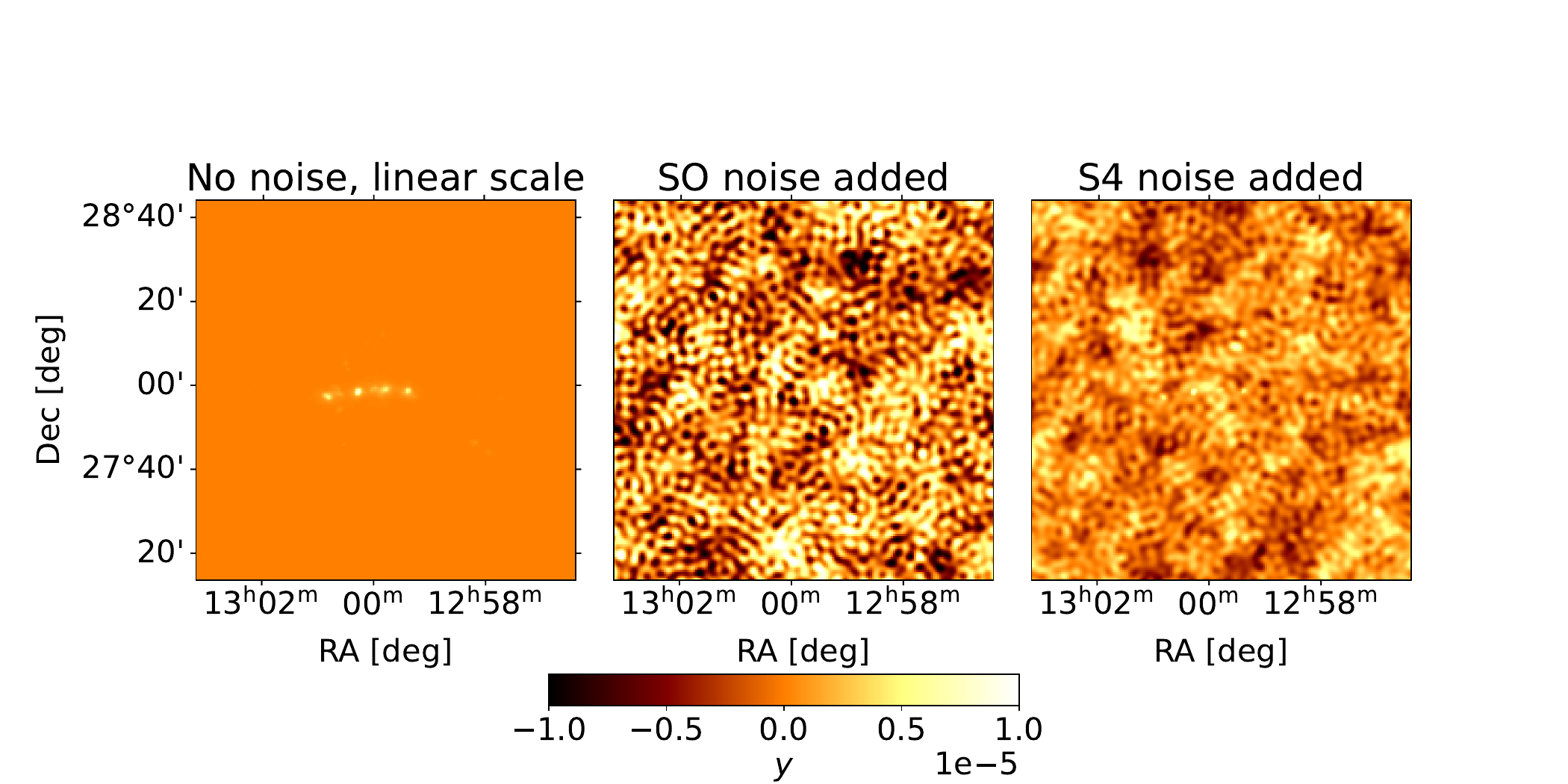}
    \caption{Left: a logarithmic-scale $y$ map of a single snapshot with high (arcsecond-scale) resolution, showing an extended structure $\sim30$~Mpc long, projected onto sky coordinates at $z=1$. The four brightest halos have masses $M\in[5.7\times10^{13},7.5\times10^{13}]\hMsun$. Right: three images of the same snapshot downgraded to a similar resolution as current and upcoming ground-based CMB experiments (0.5'/pixel) and represented with a linear colour bar centred at $y=0$. The rightmost two images contain forecasts for SO noise and CMB-S4 noise, which almost completely obscure the signal and demonstrate the need for oriented stacking.}
    \label{fig:degrading_res}
\end{figure*}

With first light expected in 2023, the SO Large Aperture Telescope will measure the CMB with an angular resolution of $\sim$1.4' at 150 GHz \citep[see][for beam sizes at all frequencies]{Zhilei2021}. This is a similarly high resolution as ACT, but will have lower noise per frequency in addition to measuring more frequencies than ACT, and thus be able to produce improved maps of the tSZ effect. CMB-S4 is a future microwave survey in the planning stages which will have comparable resolution but dramatically increased sensitivity. Figure \ref{fig:degrading_res} demonstrates a high-resolution $y$ map at resolution of 5'' in logarithmic scale, what the same $y$ map looks like in linear color scale, and mocks of the same region with forecasted SO noise and CMB-S4 noise (forecasting methods are detailed in \citealt{SO2019})\footnote{Noise models for SO are publicly available at \url{https://github.com/simonsobs/so_noise_models/tree/master/LAT_comp_sep_noise/v3.1.0} (we use the baseline tSZ model) and for S4 at \url{https://sns.ias.edu/~jch/S4_190604d_2LAT_Tpol_default_noisecurves.tgz} (we use the \texttt{yy} model with \texttt{deproj0}).}. With SO noise, none of the clusters appear visible by-eye. However, they would probably be detectable by the typical filtering that is used for SZ cluster detection in noise-dominated tSZ maps. With S4 noise, the line of four clusters is tentatively visible by-eye. However, the $y$ signal from intervening material is still below the noise. This demonstrates why oriented stacking is necessary and will continue to be necessary to detect diffuse tSZ emission, even with the next generations of CMB telescopes.

\bsp	
\label{lastpage}
\end{document}